\documentclass[useAMS,usenatbib]{mn2e}
\usepackage{graphicx,epstopdf,epsfig,bm,threeparttable,amssymb,amsmath,color}
\newcommand{\hi}{\mbox{\rm H{\sc i}}} 

\newcommand{\htwo}{\mbox{\rm H$_2$}}

\newcommand{\xcounits}{\mbox{cm$^{-2}$ (K km s$^{-1}$)$^{-1}$}}

\title[Molecular Gas Content of HI Monsters]{Molecular Gas Content of H{\Large I} Monsters and Implications to Cold Gas Content Evolution in Galaxies}
\author[Cheoljong Lee et al.]{Cheoljong Lee,$^{1,2}$ %\thanks{E-mail:cjlee@galaxy.yonsei.ac.kr}
 Aeree Chung,$^{1,3}$\thanks{E-mail: achung@yonsei.ac.kr} 
 Min S. Yun,$^{4}$ 
 Ryan Cybulski,$^{4}$
 \newauthor
 G. Narayanan,$^{4}$ 
 N. Erickson,$^{4}$ 
 \\
$^{1}$Department of Astronomy, Yonsei University, 50 Yonsei-ro, Seodaemun-gu, Seoul 120-749, Korea\\
$^{2}$Department of Astronomy, University of Virginia, Charlottesville, VA 22904, USA\\
$^{3}$Yonsei University Observatory, 50 Yonsei-ro, Seodaemun-gu, Seoul 120-749, Korea\\
$^{4}$Department of Astronomy, University of Massachusetts, 710 North Pleasant Street, Amherst, MA 01003, USA\\% \\$^{3}$National Radio Astronomy Observatory\\$^{5}$Center for Radiophysics and Space Research, Cornell University, Ithaca, NY 14853, USA\\
}
\begin{document}

\date{Accepted 0000 December 00. Received 0000 December 00; in original form 0000 October 00}

\pagerange{\pageref{firstpage}--\pageref{lastpage}} \pubyear{2012}

\maketitle

\label{firstpage}

\begin{abstract}
 We present $^{12}$CO $(J = 1 \rightarrow 0)$ observations of a sample of local galaxies ($0.04 < z < 0.08$) with a large neutral hydrogen reservoir, or ``H{\sc i} monsters''. The data were obtained using the Redshift Search Receiver on the FCRAO 14 m telescope. The sample consists of 20 H{\sc i}-massive galaxies with $M_{\rm HI} > 3\times10^{10}~M_{\odot}$ from the ALFALFA survey and 8 LSBs with a comparable $M_{\rm HI}$($>1.5\times10^{10}~M_\odot$). Our sample selection is purely based on the amount of neutral hydrogen, thereby providing a chance to study how atomic and molecular gas relate to each other in these H{\sc i}-massive systems. We have detected CO in 15 out of 20 ALFALFA selected galaxies and 4 out of 8 LSBs with molecular gas mass $M_{\rm H2}$ of ($1 - 11$)$\times10^9~M_\odot$. Their {\em total} cold gas masses of $(2 - 7)\times10^{10}~M_{\odot}$ make them some of the most gas-massive galaxies identified to date in the Local Universe. Observed trends associated with \hi, \htwo, and stellar properties of the \hi\ massive galaxies and the field comparison sample are analyzed in the context of theoretical models of galaxy cold gas content and evolution, and the importance of {\em total} gas content and improved recipes for handling spatially differentiated behaviors of disk and halo gas are identified as potential areas of improvement for the modeling.
 
\end{abstract}


\section{Introduction}

The total cold gas content of typical present day spiral galaxies like the Milky Way is known to be insufficient to maintain the current star formation rate, and additional mass accretion mechanism such as a ``cold flow" \citep{b13, b14} is needed to account for the observed properties \citep{putman06}. The gas mass fraction for such galaxies are thought to be much higher in the earlier epochs at $z \ge 1$, where the average star formation rate is also believed to be higher \citep{b5, b6, b7}. For example, the detection of extremely gas-rich galaxies both in H{\sc i} at intermediate redshift \citep{b11} and in CO at high redshift \citep{b12} indeed indicate the possibility of gas mass function evolution. This is also supported by the time evolution of co-moving $IR$ energy density in Ultra Luminous Infrared Galaxies (ULIRGs), whose burst of star formation activity requires gas-rich progenitors \citep{b9,cap07,mag11}. 

It is therefore intriguing that ALFALFA survey \citep{g05} has identified very massive H{\sc i} galaxies in nearby universe, representing local analogs with a large cold gas reservoir at higher redshift. An intriguing question to ask is {\it whether these most H{\scriptsize I}-massive galaxies are also associated with correspondingly large molecular mass and high star formation rate (SFR)}. In many numerical modeling of galaxy formation \citep[e.g.][]{dvd10}, gas accretion rate is directly translated into SFR and stellar mass build-up. However, this connection is likely more complex and is not well established observationally. Indeed, the baryon content of many H{\sc i}-massive galaxies found in the ALFALFA survey appears to be already dominated by gas and not by stars even when only atomic hydrogen is considered. 

In order to explore the connection between atomic-molecular gas and SFR further, we have conducted CO observations of a sample of galaxies at $0.04 < z < 0.08$ with a large cold gas reservoir ($M_{\rm HI} > 3\times10^{10}~M_{\odot}$), dubbed ``H{\sc i} monsters", using the Redshift Search Receiver (RSR) on the FCRAO 14 m telescope. A set of important questions to be addressed in our study are:
\begin{itemize}
\item Do the most H{\sc i}-massive galaxies also contain a large quantity of molecular gas?
\item What is the maximum cold gas (H{\sc i} $+$ H$_2$) content of normal galaxies in the field? 
\item What type of galaxies manage to accumulate such a large cold gas reservoir? 
\item What is the atomic-to-molecular gas mass ratio and is it related to other properties such as stellar mass, surface mass density, and colour?
\end{itemize}

\begin{figure*}
\label{fig_hipro}
\centering
\epsfig{file=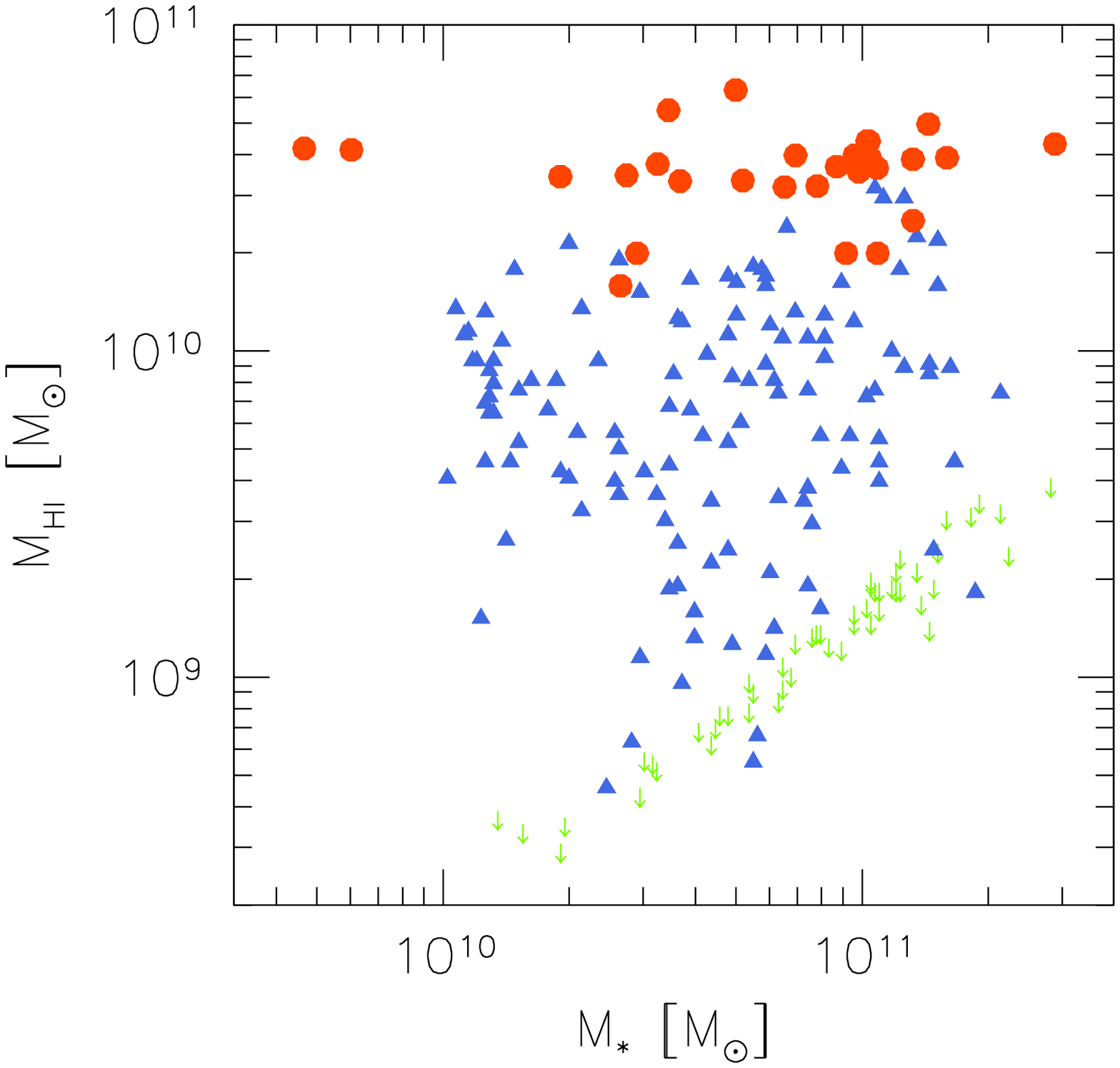,width=0.45\linewidth}\epsfig{file=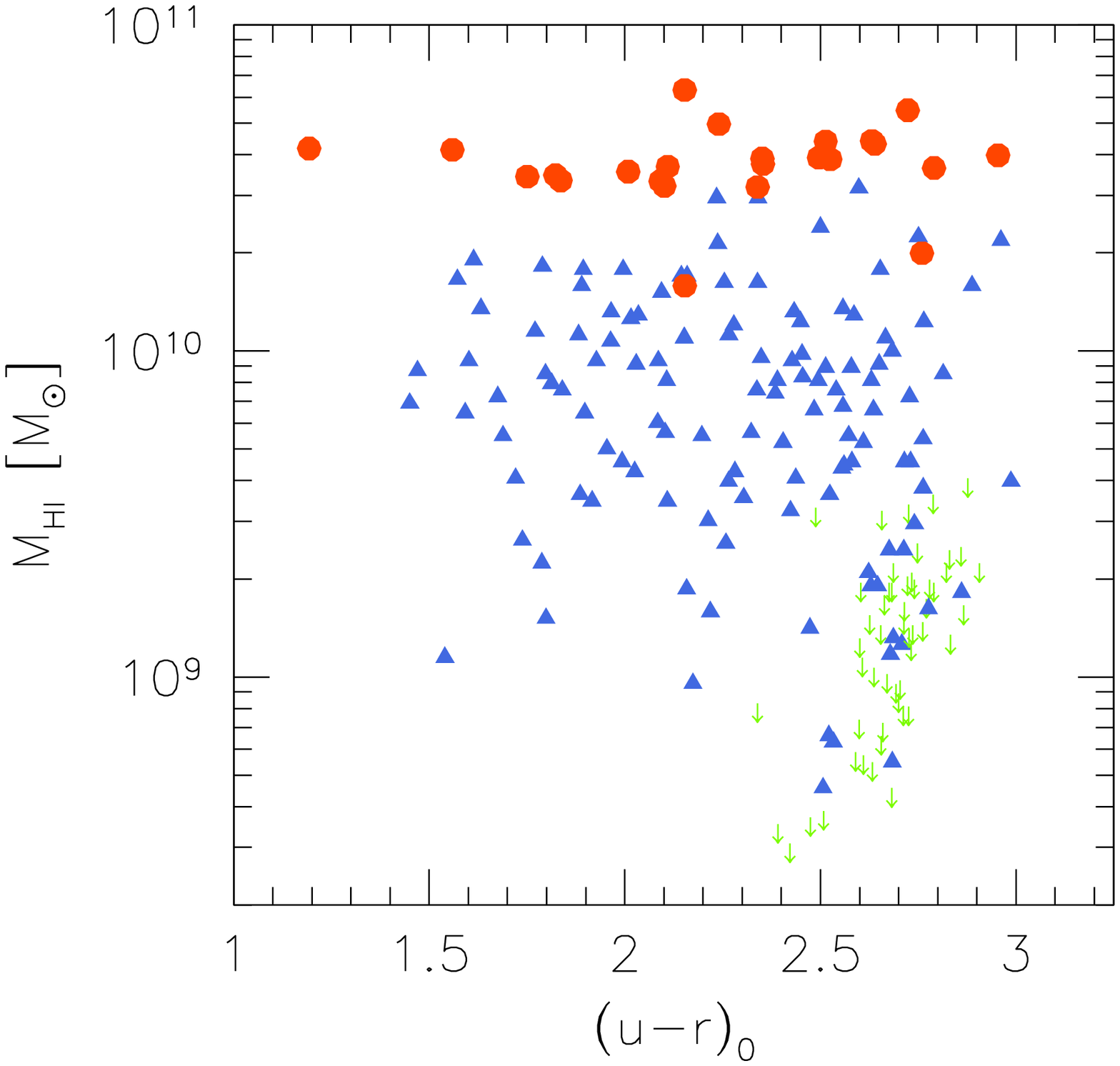,width=0.45\linewidth} 
\vspace{-0.8cm}
 \caption{Atomic gas mass ($M_{\rm HI}$) as a function of stellar mass (left) and $u$ -- $r$ colour (right) corrected for galactic extinction using the reddening maps of \citet{sch98}. Note that five galaxies among our sample whose SDSS colour is not available are missing on the right figure. H{\sc i} monsters are shown as red circles, H{\sc i} detections from COLD GASS DR2 \citep{as1, as2} as blue triangles (124 out of total 306 DR2 sample), and H{\sc i} non-detections from COLD GASS DR2 as green downward arrows (182 out of total 306 DR2 sample).Four galaxies not covered by the SDSS are not shown here.}
\end{figure*}

\begin{table*}
 \centering
  \caption{General properties of 28 H{\sc i} Monsters}
 \begin{threeparttable}
  \begin{tabular}{@{}ccccccccccc@{}}
  \hline
 ID & SDSS ID & $D_{L}$ & bt\tnote{a} & $M_\ast$ & $\mu_\ast$ & $D_{\rm 25}\tnote{a}$ & $u$ -- $r$ & $M_{\rm HI}$ & SFR$_{SDSS}$ & SFR$_{IR}$ \\
  & & [Mpc] & [mag] & [log~M$_\odot$] & [log~M$_\odot$/kpc$^2$] & [\arcsec] &  & [log~M$_\odot$] & [$M_\odot$/yr] & [$M_\odot$/yr] \\
 \hline
\multicolumn{9}{l}{- 20 ALFALFA sample}\\
AGC174522 & J075654.18$+$143827.8 & 208.41 & 16.26 & 10.71 & 8.72 & 41.51 & 1.84 & 10.52 & 5.43 & 2.43\\ 
AGC004552 & J084321.39$+$104333.8 & 205.06 & 15.07 & 11.02 & 8.48 & 58.63 & 2.35 & 10.59 & 0.69 & 1.26\\ 
AGC192542 & J090023.77$+$071828.3 & 260.12 & 17.67 &  9.67 & 7.86 & 22.81 & 1.19 & 10.62 & 3.48 & 1.15\\ 
AGC192885 & J091458.59$+$044200.7 & 244.67 & 16.05 & 11.12 & 8.91 & 49.91 & 2.53 & 10.59 & 4.58 & 4.53\\ 
AGC192040 & J094732.80$+$104508.6 & 208.23 & 17.20 & 10.54 & 9.48 & 17.30 & 2.72 & 10.74 & 0.04 & $\le2.5$\\ 
AGC005543 & J101620.49$+$044919.2 & 202.07 & 15.04 & 11.16 & 8.58 & 37.86 & 2.24 & 10.70 & 0.85 & 2.57\\ 
AGC005737 & J103353.36$+$111225.3 & 220.00 & 15.24 & 11.20 & 8.55 & 46.57 & 2.50 & 10.59 & 0.24 & 1.24\\ 
AGC205181 & J104426.09$+$152348.5 & 240.55 & 16.94 & 10.51 & 8.68 & 30.07 & 2.35 & 10.57 & 0.28 & 0.80\\ 
AGC06206A & J110949.30$+$124617.3 & 186.54 & 14.77 & 10.44 & 8.53 & 41.51 & 1.82 & 10.54 & 4.32 & 5.00\\ 
AGC215200 & J114914.07$+$153650.7 & 249.40 & 17.50 & 10.57 & 8.78 & 21.29 & 2.09 & 10.52 & 2.36 & $\le3.1$\\ 
AGC222077 & J120939.10$+$051608.6 & 194.52 & 15.60 & 10.81 & 8.55 & 49.91 & 2.34 & 10.50 & 0.43 & 0.65\\ 
AGC008585 & J133613.44$+$102841.5 & 237.02 & 15.21 & 11.46 & 8.95 & 52.26 & 2.64 & 10.64 & 0.37 & 1.29\\ 
AGC230856 & J135638.03$+$121530.6 & 253.44 & 17.02 & 10.28 & 7.94 & 42.48 & 1.75 & 10.54 & 2.68 & 1.24\\ 
AGC009162 & J141848.49$+$105037.7 & 245.82 & 15.65 & 11.02 & 8.89 & 57.30 & 2.51 & 10.64 & 0.25 & 1.47\\ 
AGC244246 & J142327.99$+$083542.1 & 247.65 & 16.67 & 11.04 & 9.23 & 36.15 & 2.79 & 10.56 & 0.11 & $\le1.2$\\ 
AGC714136 & J143407.62$+$080755.7 & 235.15 & 16.29 & 10.94 & 8.72 & 26.80 & 2.11 & 10.57 & 4.43 & 1.21\\ 
AGC009515 & J144621.37$+$130115.5 & 204.79 & 15.54 & 11.01 & 8.71 & 42.48 & 2.63 & 10.64 & 0.09 & 0.44\\ 
AGC009727\tnote{c} & J150735.63$+$143252.7 & 194.83 & 15.03 & 10.99 & 8.62 & 60.00 & 2.01 & 10.55 & ... & 2.69\\ 
AGC262058 & J160322.20$+$150029.4 & 252.52 & 17.39 &  9.78 & 7.86 & 28.72 & 1.56 & 10.62 & 2.10 & 0.48\\ 
AGC260164 & J160538.07$+$100207.4 & 176.55 & 15.20 & 10.89 & 8.62 & 48.77 & 2.10 & 10.51 & 4.31 & 3.58\\ 
\multicolumn{9}{l}{- 8 LSB sample}\\
UGC000605\tnote{b c}     & .....  & 291.46 & 16.36 & 10.98 &  ... & 52.26 & ...  & 10.60 & ... & 2.09\\ 
UGC001941\tnote{b c}     & .....  & 200.03 & 16.54 & 11.12 &  ... & 80.94 & ...  & 10.40 & ... & 0.64 \\ 
UGC002224\tnote{b c}     & .....  & 162.08 & 16.84 & 10.96 &  ... & 53.48 & ...  & 10.30 & ... & 0.41\\ 
PGC089535\tnote{b c}     & .....  & 194.88 & 16.71 & 10.46 &  ... & 47.66 & ...  & 10.30 & ... & 0.50\\ 
UGC005440 & J100535.79$+$041645.8 & 280.45 & 16.96 & 10.70 & 8.35 & 49.91 & 2.15 & 10.80 & 2.14 & 0.91\\ 
UGC006124 & J110339.49$+$315129.3 & 203.92 & 16.36 & 11.04 & 9.23 & 72.14 & 2.76 & 10.30 & 0.24 & 1.00\\ 
PGC089614 & J123036.25$+$243448.7 & 297.37 & 17.56 & 10.42 & 8.31 & 32.22 & 2.15 & 10.20 & 1.89 & $\le3.7$\\ 
PGC042102 & J123659.34$+$141949.3 & 370.95 & 18.27 & 10.84 & 9.04 & 11.97 & 2.95 & 10.60 & 0.08  & $\le7.7$\\ 
 \hline
\end{tabular}
 \begin{tablenotes}

\item[a] Data compiled from Hyper-Leda.
\item[b] SDSS data not available.
\item[c] Stellar masses are computed from K band magnitude.
\end{tablenotes} 
\end{threeparttable} 
\label{table_properties}
\end{table*}

In this work, we complement our atomic and molecular gas mass data with available SDSS photometry (DR7; \citealt{sdssdr7}) and MPA-JHU DR7\footnote{http://www.mpa-garching.mpg.de/SDSS/DR7/} catalogue (\citealt{kauffmann03}) to extract information of the sample galaxies such as stellar mass, stellar surface mass density and colour. We also analyze the infrared-derived star formation rate using the 22 \micron\ photometry from the Wide-field Infrared Survey Explorer \citep[WISE; ][]{wright10} in order to account for dust obscured star formation activity.  With this information, we compare H{\sc i} monsters with galaxies having similar stellar mass range from COLD GASS survey \citep{as1,as2}, which we adopt as a { \color{black}reference sample to examine whether our H{\small I} monsters are exceptional in their properties}.

This paper is organized in the following order. In \S\ref{sample}, we describe our sample selection criteria and general properties of the sample. In \S\ref{obs and data reduction} we describe the observations and data reduction. In \S\ref{results} we present atomic and molecular properties of H{\sc i} monsters compared to the { reference sample}, and discuss our results in \S\ref{discussion}. We summarize our results and discussions in \S\ref{summary}. Throughout this paper, we assume a standard flat $\Lambda$CDM cosmology with $H_0$ = 70 km s$^{-1}$ Mpc$^{-1}$, and we adopt a constant ${\rm CO} - {\rm H_{2}}$ conversion factor of $\alpha_{\rm CO}$ = $3.2~M_\odot$~(K~km~s$^{-1}$ pc$^2$)$^{-1}$, unless otherwise mentioned. The conversion factor we use corresponds to $X_{\rm CO} = 2\times10^{20}$~\xcounits, which is commonly accepted value for MW like normal spirals (\citealt{sm96}). Note that we do not include helium in the calculation in this work.

\section{Sample}
\label{sample}
\subsection{H{\small I} monsters}
\label{monsters}

Our sample primarily consists of the most H{\sc i}-massive galaxies ($M_{\rm HI}>3\times10^{10}~M_\odot$) within the redshift range, $0.04 < z < 0.08$ in the ALFALFA catalog as of March 2008 (\citealt{haynes11}). The lower \textit{z} limit was imposed by the receiver upper bound frequency limit at 111 GHz, which corresponds to \textit{z} of $\sim$ 0.037 in $^{12}$CO $(J = 1 \rightarrow 0)$ frequency ($\nu_{\rm rest}$=115.271 GHz). The upper \textit{z} limit of our sample represents the depth of the H{\sc i} survey in Spring, 2008. We also included 8 additional galaxies in the similar redshift range $(0.04 < z < 0.083)$ with $M_{\rm HI}>10^{10}~M_\odot$ identified in the H{\sc i} survey of low surface brightness galaxies \citep[LSBs][in preparation]{oneil14}. The general properties of 28 galaxies in the sample are summarized in Table~\ref{table_properties}. Stellar masses in this table are collected from the MPA-JHU DR7 catalogue, or converted from $K$-band magnitude if the galaxy is not catalogued. Our targets are comparable with the sample of COLD GASS survey in stellar mass and colour, which makes a direct comparison of optical and gas properties straightforward.
%Evidently, most galaxies satisfy the selection criterion of COLD GASS survey ($M_{\ast} > 10^{10}~M_{\odot}$), which makes a direct comparison of optical and gas properties straightforward. 

\begin{figure*}
\includegraphics[width=1\linewidth]{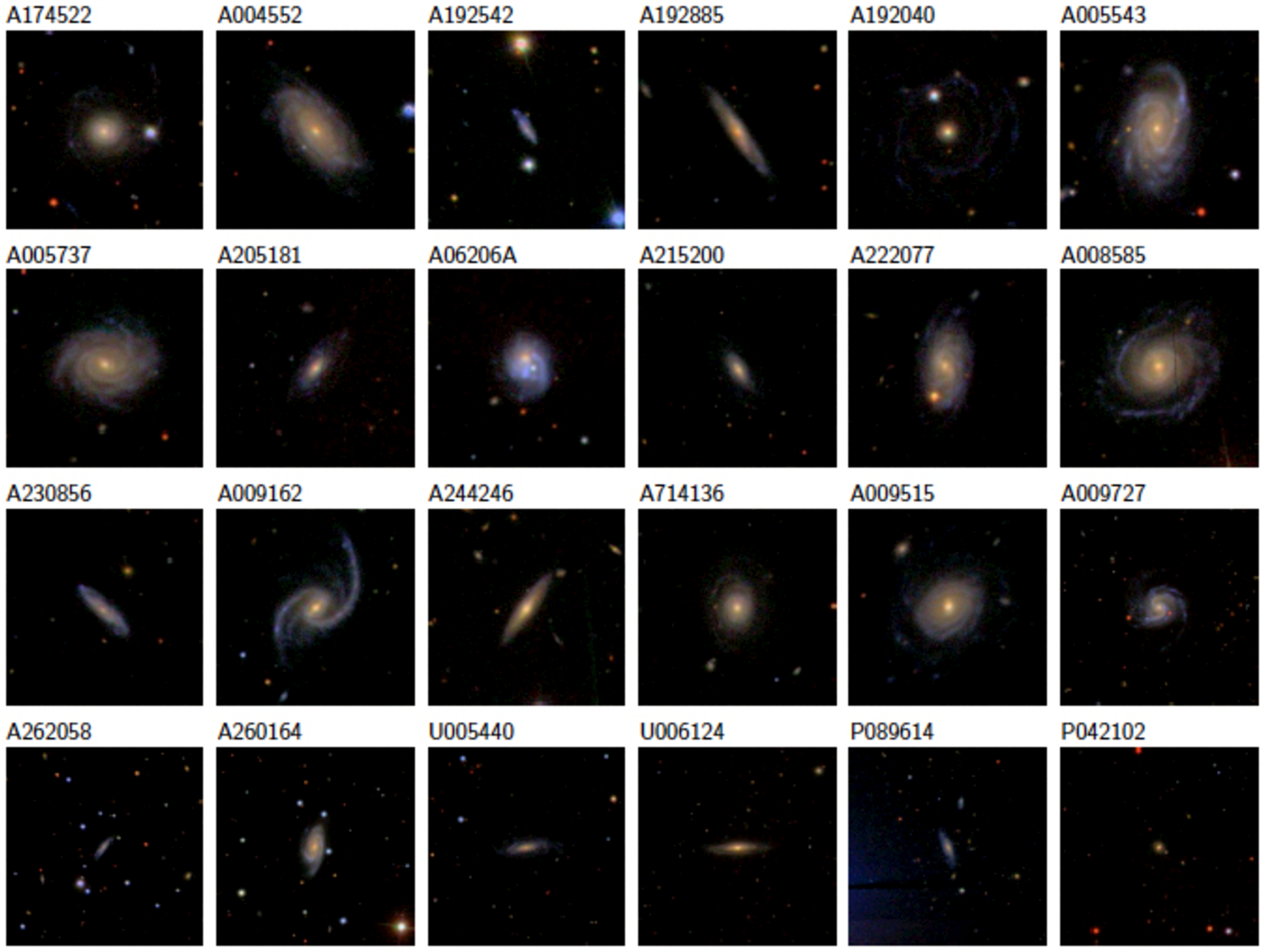}
 \caption{Available SDSS optical images of 24 H{\sc i} monsters ($1.8\times1.8$ arcmin$^2$). The majority of them are optically normal spirals (Sb - Sd) with blue, faint disks. A number of galaxies show potential signatures of tidal interactions (e.g. AGC174522, AGC192040, AGC06206A, AGC009162 and AGC009515).   \label{fig_sdssim}}
\end{figure*}

By the sample definition, the most notable feature of our sample galaxies is their massive H{\sc i} disks.  In Figure~\ref{fig_hipro}, we compare the atomic gas mass ($M_{\rm HI}$) of H{\sc i} monsters with the { reference sample}, compiled from COLD GASS DR2 (306 galaxies). In this figure, those with H{\sc i}-detections (124/306) and H{\sc i}-upper limits (182/306) in COLD GASS galaxies are shown in blue triangles and green down-arrows, respectively. Our H{\sc i} monsters are shown in red circles. The atomic gas mass of H{\sc i} monsters ranges from $1.6\times10^{10}~M_\odot$ (PGC 89614) to $6.3\times10^{10}~M_\odot$ (UGC 05440), which are at the high mass end of COLD GASS at given stellar mass or $u$ -- $r$ colour.  

{ Note} that {\em our H{\small I} monster sample selected for their high gas content alone span the entire range of stellar mass and optical colour nearly uniformly}, unlike in many other sample selections where everything tends to scale with everything else. The H{\sc i} gas mass ratio of the selected galaxies ranges quite widely with $0.15<M_{\rm HI}/M_*<8.9$, and our H{\sc i} monster sample also includes normal galaxies when H{\sc i} mass is scaled by stellar mass or optical luminosity. Except for a few cases however, $M_{\rm HI}/L_B$ of our sample is comparable or more than that of Sd population \citep[Fig. 4 of][]{rh94}, and most galaxies in the sample are not only gas-massive but also gas-rich.

The Sloan Digital Sky Survey colour images shown in Figure~\ref{fig_sdssim} suggest that most H{\sc i} monsters are normal spirals (Sb - Sd) while some objects show morphological peculiarities (e.g. AGC174522 and AGC192040 with a very faint and extended disk; AGC06206A with an unusually blue and asymmetric stellar disk).  In most cases, no obvious bright companion is seen within a $\sim2'$ radius.

\subsection{COLD GASS : the {\color{black}reference} sample}
\label{coldgass}
Recently, a series of ongoing studies - the GALEX Arecibo SDSS Survey (GASS : \citealt{c10}) and the CO Legacy Database for GASS (COLD GASS : \citealt{as1, as2}) - are trying to establish a global gas scaling relation in $\sim 300$ local massive galaxies ($M_\ast > 10^{10}~M_{\odot}$). By combining high quality Arecibo H{\sc i} data and IRAM CO data with SDSS photometry, spectroscopy and GALEX imaging, they provide an unbiased view on the mean atomic and molecular gas properties of galaxies covering a large dynamic range. 

In this work, we employ the results from these homogeneous H{\sc i} and CO data as a reference when we study gas content of massive galaxies. Details of the COLD GASS survey and its sample selection are described by \citet{as1}.  Here we offer only a brief summary: 
\begin{enumerate}
\item COLD GASS sample is a subset of the GALEX Arecibo SDSS Survey (GASS) galaxies, which consist of $\sim1000$ local massive galaxies ($M_\ast > 10^{10}~M_{\odot}$) randomly selected from a larger parent sample of galaxies in the redshift range of $0.025 < z < 0.05$ within the region covered by the SDSS, the ALFALFA and the GALEX survey. 
\item GASS galaxies are selected to have a flat $M_\ast$ distribution. %, and therefore COLD GASS galaxies do so as well.
\item Among $\sim1000$ COLD GASS galaxies, $\sim$ 350 targets are selected randomly as COLD GASS sample. We only use COLD GASS DR2 sample (total 306 galaxies), published in \citet{k12}.
\item The COLD GASS galaxies have been integrated until the CO line was detected, or an upper limit on the molecular gas mass fraction ($M_{\rm H2}/M_{\ast}$) of $\sim$ 1.5\% was reached. 
\item Most COLD GASS galaxies were observed only at the central position, and an aperture correction is applied to estimate the total molecular gas content \citep[see][]{as1}.
\end{enumerate}
{ \color{black}Although the COLD GASS galaxies have been selected by their stellar mass while our H{\small I} monsters have been selected based on their H{\small I} mass, both samples are found to have a similar stellar mass range, except for two exceptionally low mass galaxies in our sample. In addition, both samples cover a comparable fraction of galaxies in mass bins, similarly distributed from low to high stellar mass. Hence our comparison between the two samples allow us to study how H{\small I} massive otherwise normal galaxies behave in total gas and star formation properties. } 

\subsection{Description of quantities}
\label{quantities}
Here, we explain how we compute the quantities appearing in this work.
\begin{enumerate}
\item Stellar mass ($M_{\ast}$), SFR$_{SDSS}$ are compiled from MPA-JHU DR7 catalogue, if available. For the galaxies with no available stellar masses, we compute them from the 2MASS \textit{K}$-$band magnitude of the galaxies using a linear relation between the \textit{K}$-$band magnitude and stellar mass, fitted from the galaxies with available stellar masses from MPA-JHU catalogue.
\item Stellar surface mass density ($\mu_\ast$) is calculated (also following the MPA-JHU DR7 catalog) as 
\begin{equation}
\label{eq_mu}
\mu_\ast = \frac{M_{\ast}}{2\pi {R}^{2}_{50,z}},
\end{equation} 
where  ${R}_{50,z}$ is the SDSS \textit{z}$-$band 50\% flux intensity petrosian radius in kiloparsecs.
\item Star formation rate derived from $IR$ luminosity (SFR$_{IR}$) is computed from the Wide-field Infrared Survey Explorer \citep[WISE][]{wright10} 22 \micron\ band photometry, adopting the calibration (Eq.~5) by \citet{m11}.  
\end{enumerate}

\section{Observations and Data Reduction}
\label{obs and data reduction}
The observations were conducted using the FCRAO 14~m telescope in April and May 2008. The 14 m telescope is radome-enclosed single-dish millimeter telescope, located in Massachusetts. The spectra were taken using the redshift search receiver (RSR), a sensitive, ultra-wideband spectrometer that was developed at the University of Massachusetts. The receiver is one of the facility instruments built for the 50~m diameter Large Millimeter Telescope (LMT). While the LMT was under its final construction, the RSR was installed on the 14 m and used for several science projects from 2006 to 2009.

The RSR consists of a set of wide-band analog autocorrelation spectrometers. There are four receivers each covering 74-111 GHz simultaneously in a dual-beam, dual polarization with a spectral resolution of 31 MHz. Mechanically six spectrometers, covering 6.5 GHz each, are connected to cover a total bandwidth of 37 GHz with each receiver. See \citet{e07} and \citet{c09} for further details about the RSR system. In 2008, 12 spectrometers among 24 (4 receivers and 6 spectrometers) were available, and we had all four receivers cover 92-111 GHz using three spectrometers. In this paper, we discuss only the data from the spectrometer covering the highest frequency band (103 - 111 GHz) where $^{12}$CO $J = 1 \rightarrow 0$ line of our sample $(0.04 < z < 0.08)$ appears.  The diffraction limited beam size at 115 GHz is 50\arcsec, which is large enough to cover the entire stellar disks of target galaxies in nearly all cases (see Fig.~\ref{fig_rsrsp}).

The on-source integration time  ranges between 0.8 and 9.3 hr (a median of 2.0 hr) on each source depending on the weather and the CO line strength (Table~\ref{table_properties}). The typical system temperature during the observing season was $230\pm25$ K. The calibration and the focus were done in the same manner as for the local ULIRG study of \citet{c09}, and details can be found in the reference. This yields a typical rms sensitivity of $\sigma\approx 0.24-0.61$ mK in $T_A^*$ (corrected antenna temperature for the atmosphere absorption and spillover losses) with the mean of $0.38\pm0.08$ mK.  We adopt the antenna gain G of  45  Jy~K$^{-1}$ in $T_A^*$ which has been determined by monitoring the intensity of planets.

The data have been reduced using the SPAPY, a data reduction software that has been developed by G. Narayanan for the RSR spectrum analysis. In order to obtain the final spectrum of each galaxy, noisy data due to bad weather or instrumental failures were first excluded by visual inspection.
A linear baseline was fit over the continuum channels to calculate the rms noise of each scan, then the scans were averaged weighted by noise. For most objects, a linear baseline was removed while a second order baseline was removed in some cases. The rms noise of the reduced spectra of the sample is listed in Table~\ref{table_properties}. Also, the final RSR spectra along with DSS images of H{\sc i} monsters are presented in Figure~\ref{fig_rsrsp}.

Features showing a $>3\sigma$ bump within a few 100~km~s$^{-1}$ around the inferred CO frequency from the optical redshift are defined as detections.

\begin{figure*}
\centering
\epsfig{file=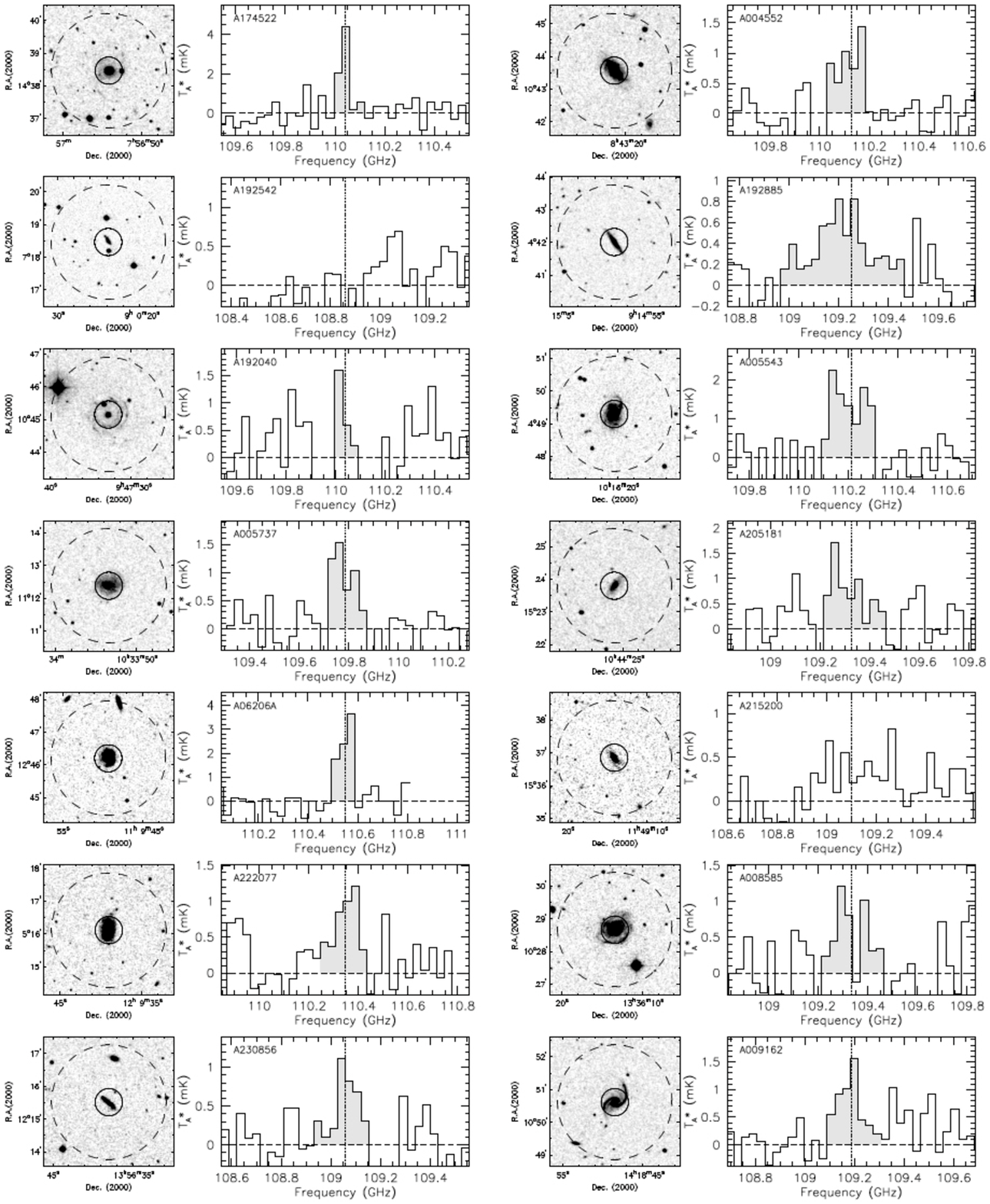,width=0.95\linewidth,clip=} 
 \caption{4' x 4' DSS images (left panel) and final RSR spectra (right panel) of H{\sc i} monsters. In DSS images, dotted circles denote the size of Arecibo beam (3.5') and filled circles indicate RSR beam size (50"). Contrast of galaxy images are adjusted to show faint features (high contrast). On the right panel, the RSR spectra are presented in $T_A^*$. Only 1 GHz (out of $\approx 6.5~$GHz bandwidth in total) around the CO frequency is shown. The shaded channels indicate CO emissions for detected galaxies. For non detections, 1 GHz around the expected CO frequency for given optical redshift is shown without shading. The dotted line in each spectrum represents the measured or the expected CO frequency. Each spectrum is scaled by the peak in $T_A^*$, showing the range of -0.2 to 1.25$\times$peak (-0.2 to 1.25$\times3\sigma$ for non-detections).}
\label{fig_rsrsp}
\end{figure*}

\setcounter{figure}{2}
\begin{figure*}
\centering
\epsfig{file=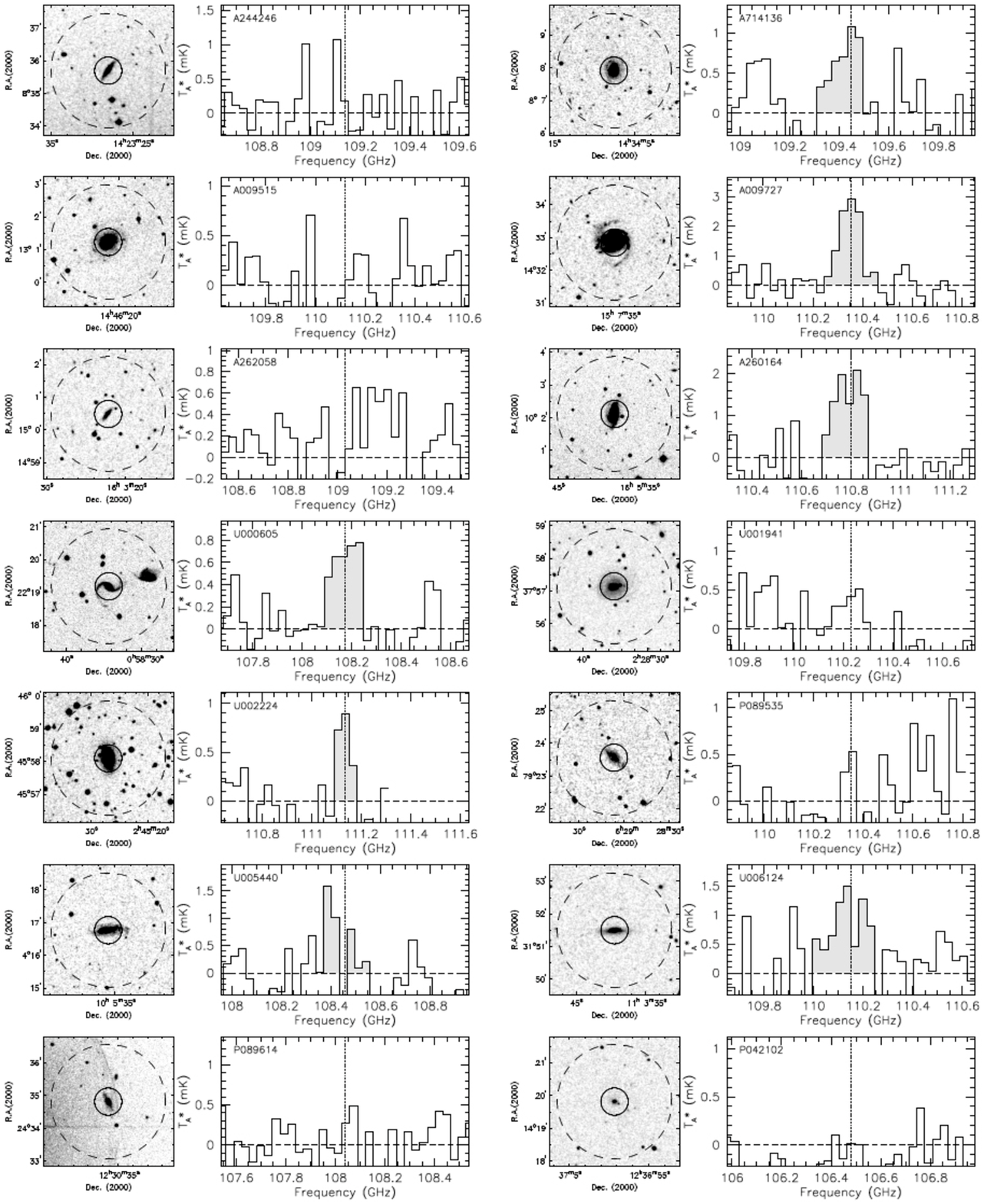,width=0.95\linewidth,clip=} 
\caption{(Continued)}
\end{figure*}

\begin{table*}
 \centering
  \caption{Molecular gas masses and CO quantities for the H{\sc i} Monsters}
 \begin{threeparttable}
  \begin{tabular}{@{}ccccccccc@{}}
 \hline
ID & $\nu_{\rm obs}$ & rms & $t_{\rm int}$ & $I_{\rm CO}$\tnote{a} & {$W_{\rm CO}$\tnote{b}} &  $L_{\rm CO} \tnote{a} $ & $M_{\rm H2}$\tnote{a} & {Note\tnote{c}}\\	
  & [GHz] & [mK] & [hrs] &[K~km~s$^{-1}$] & {[km~s$^{-1}$]} &  [K~km~s$^{-2}$~pc$^2$] & [log~M$_\odot$]  & \\	
 \hline	
AGC174522 & 110.04 & 0.34 & 2.20 & 0.59 $\pm$ 0.06 & 134.55 $\pm$ 18.03 & 9.44 & 9.95 & {\sc ii}\\
AGC004552 & 110.13 & 0.20 & 4.12 & 0.34 $\pm$ 0.04 & 238.11 $\pm$ 45.07 & 9.19 & 9.69 & {\sc i}\\
AGC192542 & 108.86 & 0.37 & 0.82 & $<$0.30 & 324.00&  $<$9.34 & $<$9.84 & ...\\
AGC192885 & 109.25 & 0.23 & 3.51 & 0.44 $\pm$ 0.08 & 536.92 $\pm$ 178.56 & 9.45 & 9.95 & {\sc iii}\\
AGC192040 & 110.04 & 0.43 & 1.80 & 0.16 $\pm$ 0.07 & 98.34 $\pm$ 53.86 & 8.86 & 9.37 & {\sc ii}\\
AGC005543 & 110.21 & 0.39 & 2.41 & 0.77 $\pm$ 0.09 & 342.20 $\pm$ 71.90 & 9.53 & 10.03 & {\sc i}\\
AGC005737 & 109.79 & 0.27 & 1.39 & 0.37 $\pm$ 0.06 & 242.29 $\pm$ 56.26 & 9.28 & 9.79 & {\sc i}\\
AGC205181 & 109.33 & 0.41 & 1.04 & 0.45 $\pm$ 0.11 & 259.53 $\pm$ 88.90 & 9.44 & 9.94 & {\sc iii}\\
AGC06206A & 110.55 & 0.44 & 1.35 & 0.65 $\pm$ 0.08 & 179.70 $\pm$ 30.14 & 9.39 & 9.89 & {\sc ii}\\
AGC215200 & 109.10 & 0.33 & 1.67 & $<$0.38 & 377.00&  $<$9.41 & $<$9.91 & ...\\
AGC222077 & 110.35 & 0.40 & 1.98 & 0.35 $\pm$ 0.09 & 293.65 $\pm$ 122.48 & 9.16 & 9.67 & {\sc i}\\
AGC008585 & 109.33 & 0.36 & 3.80 & 0.36 $\pm$ 0.15 & 300.79 $\pm$ 194.48 & 9.35 & 9.85 & {\sc i}\\
AGC230856 & 109.06 & 0.35 & 2.69 & 0.30 $\pm$ 0.09 & 272.52 $\pm$ 114.76 & 9.31 & 9.82 & {\sc i}\\
AGC009162 & 109.19 & 0.36 & 1.94 & 0.41 $\pm$ 0.09 & 262.62 $\pm$ 82.68 & 9.42 & 9.93 & {\sc ii}\\
AGC244246 & 109.14 & 0.42 & 2.88 & $<$0.77 & 581.00&  $<$9.70 & $<$10.21 & ...\\
AGC714136 & 109.45 & 0.33 & 2.59 & 0.29 $\pm$ 0.08 & 271.46 $\pm$ 110.46 & 9.23 & 9.74 & {\sc iii}\\
AGC009515 & 110.12 & 0.29 & 2.02 & $<$0.40 & 382.00&  $<$9.26 & $<$9.77 & ...\\
AGC009727 & 110.35 & 0.36 & 2.08 & 0.84 $\pm$ 0.09 & 287.54 $\pm$ 46.59 & 9.54 & 10.04 & {\sc i}\\
AGC262058 & 109.03 & 0.27 & 6.12 & $<$0.44 & 508.00&  $<$9.48 & $<$9.98 & ...\\
AGC260164 & 110.80 & 0.56 & 1.63 & 0.67 $\pm$ 0.13 & 324.74 $\pm$ 107.62 & 9.35 & 9.86 & {\sc i}\\
UGC000605 & 108.17 & 0.27 & 4.02 & 0.29 $\pm$ 0.06 & 370.28 $\pm$ 147.79 & 9.41 & 9.92 & {\sc ii}\\
UGC001941 & 110.23 & 0.37 & 9.29 & $<$0.43 & 383.00&  $<$9.27 & $<$9.77 & ...\\
UGC002224 & 111.14 & 0.32 & 5.94 & 0.15 $\pm$ 0.05 & 173.70 $\pm$ 86.54 & 8.64 & 9.15 & {\sc iii}\\
PGC089535 & 110.35 & 0.31 & 6.49 & $<$0.47 & 424.00&  $<$9.29 & $<$9.79 & ...\\
UGC005440 & 108.46 & 0.54 & 1.98 & 0.18 $\pm$ 0.12 & 112.58 $\pm$ 82.73 & 9.17 & 9.67 & {\sc i}\\
UGC006124 & 110.15 & 0.48 & 1.31 & 0.53 $\pm$ 0.12 & 352.13 $\pm$ 140.31 & 9.37 & 9.88 & {\sc iii}\\
PGC089614 & 108.04 & 0.36 & 1.65 & $<$0.44 & 412.00&  $<$9.62 & $<$10.12 & ...\\
PGC042102 & 106.48 & 0.30 & 1.98 & $<$0.27 & 355.00&  $<$9.58 & $<$10.09 & ...\\
\hline
\end{tabular}
 \begin{tablenotes}
\item[a] For CO non-detections, 3$\sigma$ upper limits are estimated.
\item[b] For CO non-detections, H{\sc i} line widths are presented.
\item[c] Optical morphology class. Details can be found in \S\ref{results}.1.
\end{tablenotes} 
\end{threeparttable}
 \label{table_co}
\end{table*}

\section{Results}
\label{results}

Among the 28 galaxies observed, 15 (out of 20) ALFALFA selected galaxies and 4 (out of 8) LSB galaxies are detected in CO with a detection rate of 75\% and 50\%, respectively.  Detailed descriptions of the observational results and derived quantities are presented in this section.  %In the following subsections, we give a result of RSR observations of H{\sc i} monsters.

\subsection{Measurement of the CO quantities and ${\bm M}_{\bf H2}$}
\label{CO measurement}

The integrated intensity $(I_{\rm CO})$, central frequency $(\nu_{\rm CO})$ and CO line width $(W_{\rm CO})$ are determined from the final reduced spectra. We calculate the CO line luminosity following \citet{s92},
\begin{equation}
  L_{\rm CO}^{'}=3.25\times10^7 I_{\rm CO}~\nu_{\rm obs}^{-2}~ D_{L}^2~(1+\textit{z})^{-3}~\rm{K~km~s^{-1}~pc^2}
\end{equation}
where $I_{\rm CO}$ is the velocity integrated CO intensity in Jy~km~s$^{-1}$, $D_{L}$ is the luminosity distance in Mpc, $\nu_{\rm obs}$ is the observed frequency in GHz, and \textit{z} is the redshift of the object. We use the redshift determined from the CO line to calculate $D_{ L}$. For undetected sources, 3$\sigma$ upper limits are computed using their H{\sc i} line widths and optical redshifts. 

{\em The first important result revealed by this study is that these H{\small I} monsters are also extremely massive in molecular gas as well.}
The derived CO luminosity of the detected sample is $(4.4 - 34.6)\times10^{8}~$K~km~s$^{-1}$~pc$^2$, corresponding to H$_2$ masses of $(1.4 - 11.1)\times10^9~M_{\odot}$ using a constant Galactic ${\rm CO} - {\rm H_{2}}$ conversion factor of $\alpha_{\rm CO}$ = $3.2~M_\odot$~(K~km~s$^{-1}$ pc$^2$)$^{-1}$ (same as the COLD GASS sample; \citealt{sm96}). The median value of H$_2$ masses is $7.2\times10^9~M_{\odot}$, which is 50\% larger than that of the $L^*$ galaxy of the local CO luminosity function \citep[$\approx4.7\pm0.9\times10^9~M_\odot$;][adopting the same conversion factor]{k03,bwl13}. Therefore, H{\sc i} monsters represent some of the most molecular gas-massive objects in the Local Universe.

The CO-to-H$_2$ conversion factor is known to have some dependence on metallicity, particularly in low metallicity environment \citep[e.g.][]{b02,i97,t98}. However, the metallicity of our sample galaxies derived using the SDSS spectra and Eq. (7) of \citet{kd02} is 12$+$log(O/H) $\gtrsim8.6$ in most cases, as expected from their stellar masses.  Therefore, these molecular gas masses $M_{\rm H2}$ derived using the standard conversion factor should be reliable.  
The non-detections of the two least massive galaxies in our sample (AGC192542 and AGC262058) may be the result of this metallicity dependence, and it is possible that these two galaxies still harbor some molecular hydrogen gas untraced by CO. We summarize the CO related quantities and the molecular gas masses in Table~\ref{table_co}. 

\begin{figure*}
\centering
\epsfig{file=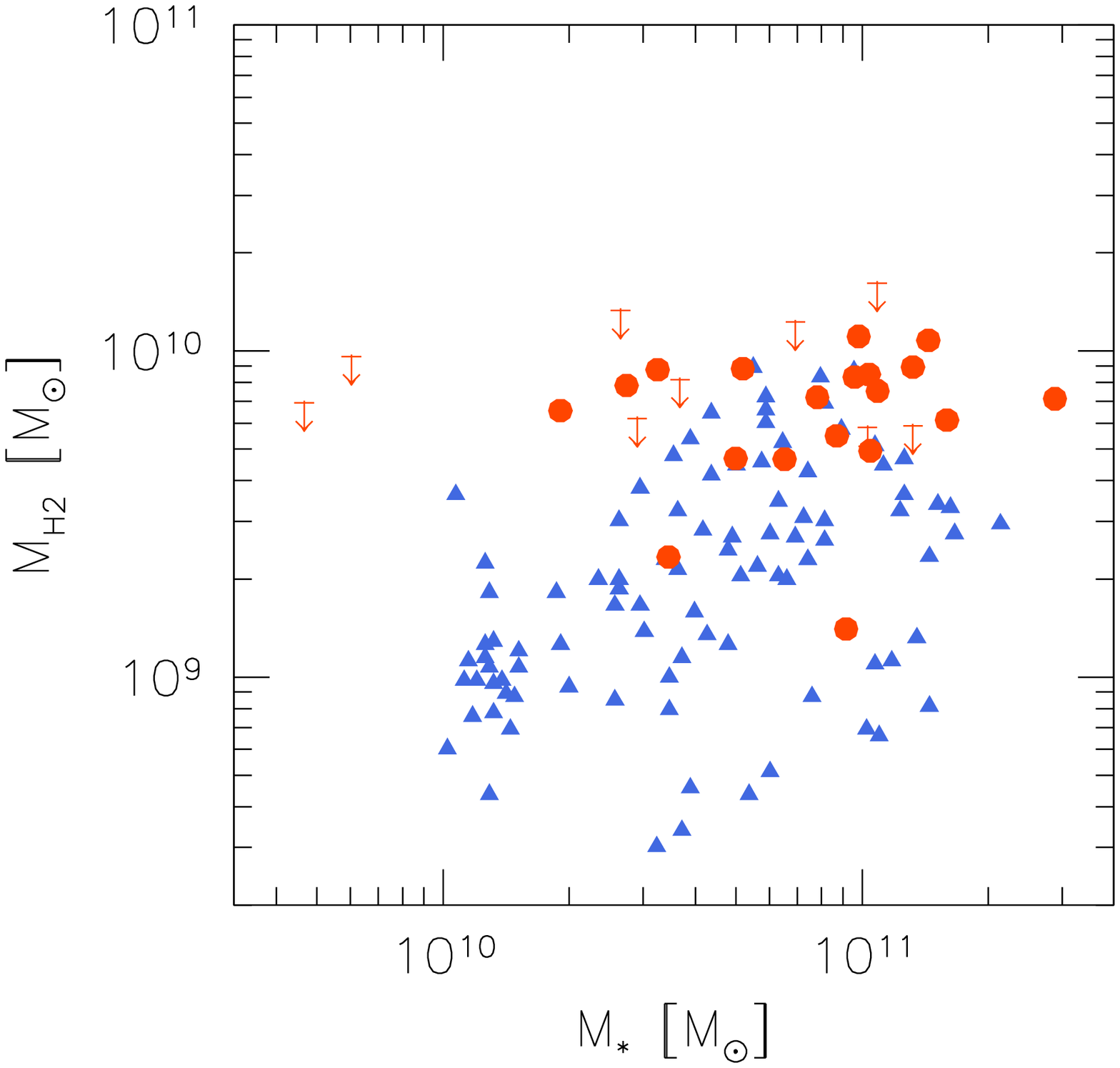,width=0.45\linewidth,clip=}\epsfig{file=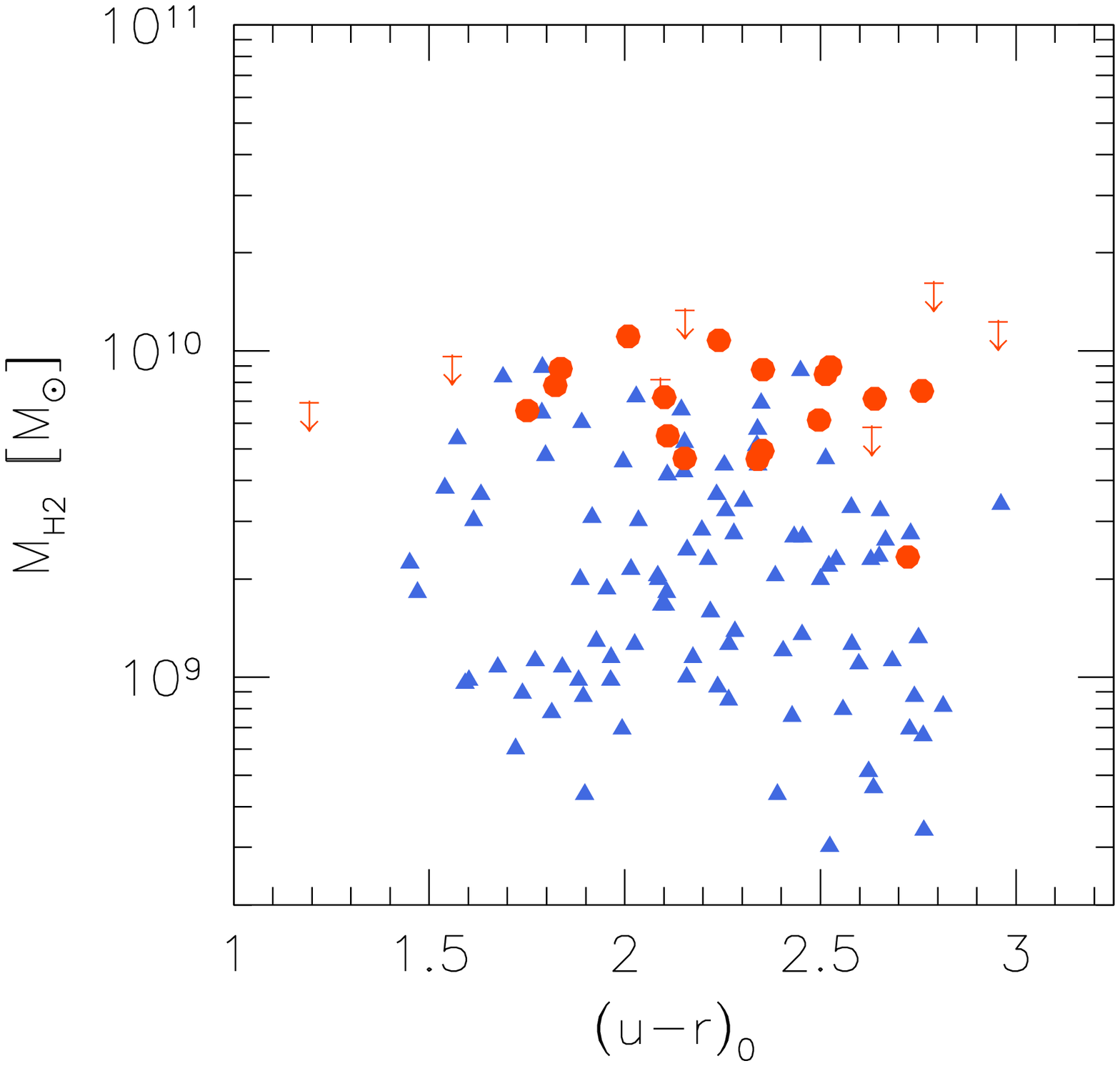,width=0.45\linewidth,clip=} 
\vspace{-0.8cm}
 \caption{Molecular gas mass ($M_{\rm H2}$) as a function of stellar mass (left) and $u-r$ colour (right) corrected for galactic extinction using the reddening maps of \citet{sch98}. CO detected H{\sc i} monsters are indicated by red circles and 3$\sigma$ upper limits for non-detections are shown by down-arrow. Blue triangles represent COLD GASS galaxies that are detected both in H{\sc i} and CO (88 out of total 306 DR2 sample).}
\label{fig_h2pro}
\end{figure*}

{\em Another key result is that the CO-detected H{\small I} monsters also span a wide range in stellar mass and optical colour, nearly the full range of values associated with the field comparison sample.}  Molecular gas masses of the H{\sc i} monsters are plotted as a function of stellar mass and colour in Figure~\ref{fig_h2pro} along with those of the CO-detected COLD GASS galaxies for comparison.  The two lowest stellar mass galaxies with the bluest $u-r$ colour are undetected, but otherwise the CO-detected \hi\ monsters span the same range of stellar mass and colour.   The H{\sc i} monsters are found at the high end of molecular gas mass distribution in most cases, supporting the idea that galaxies with  a large neutral atomic hydrogen reservoir can also support a large molecular gas disk.  The dispersion in the molecular gas content is larger than in the \hi, and the dependencies of molecular gas content (CO luminosity) on H{\sc i}, optical and SFR properties are discussed further in the following sections.  %On the other hand, the CO-detected HI monsters span the full range of $M_*$ and $u$ -- $r$ colour, and no specific optical (stellar) properties are strongly correlated with the molecular gas mass. %However, the monsters are quite widely spread in $M_{\rm H2}$, and not necessarily the richest in molecular gas. %In addition, CO-detected monsters are not systematically more luminous in the visible (e.g. $K$ and $B$) or optically bluer. 

Optical morphology of the CO-detected H{\sc i} monsters can be roughly divided into three categories as noted in Table~\ref{table_co}: {\sc(i)} a group with typical spiral arms (e.g. AGC005737), {\sc (ii)} a group showing tidal features (e.g. AGC192040) and lastly, {\sc(iii)} a group with red colour or low surface brightness disk (e.g. UGC006124). One exception of the second class is AGC09515, which is distinct from the others in the same group by showing a number of small stellar complexes around it, indicating the accretion of dwarf satellites rather than a major merging event, which is not always expected to contain much molecular gas. Most galaxies with red colour and/or not well-defined spiral structure (e.g. PGC042102) are not detected in CO.  

\subsection{The relationship between atomic and molecular gas among H{\small I} Monsters}
\label{relation_h1_h2}
\begin{figure*}
\centering
\centering
\epsfig{file=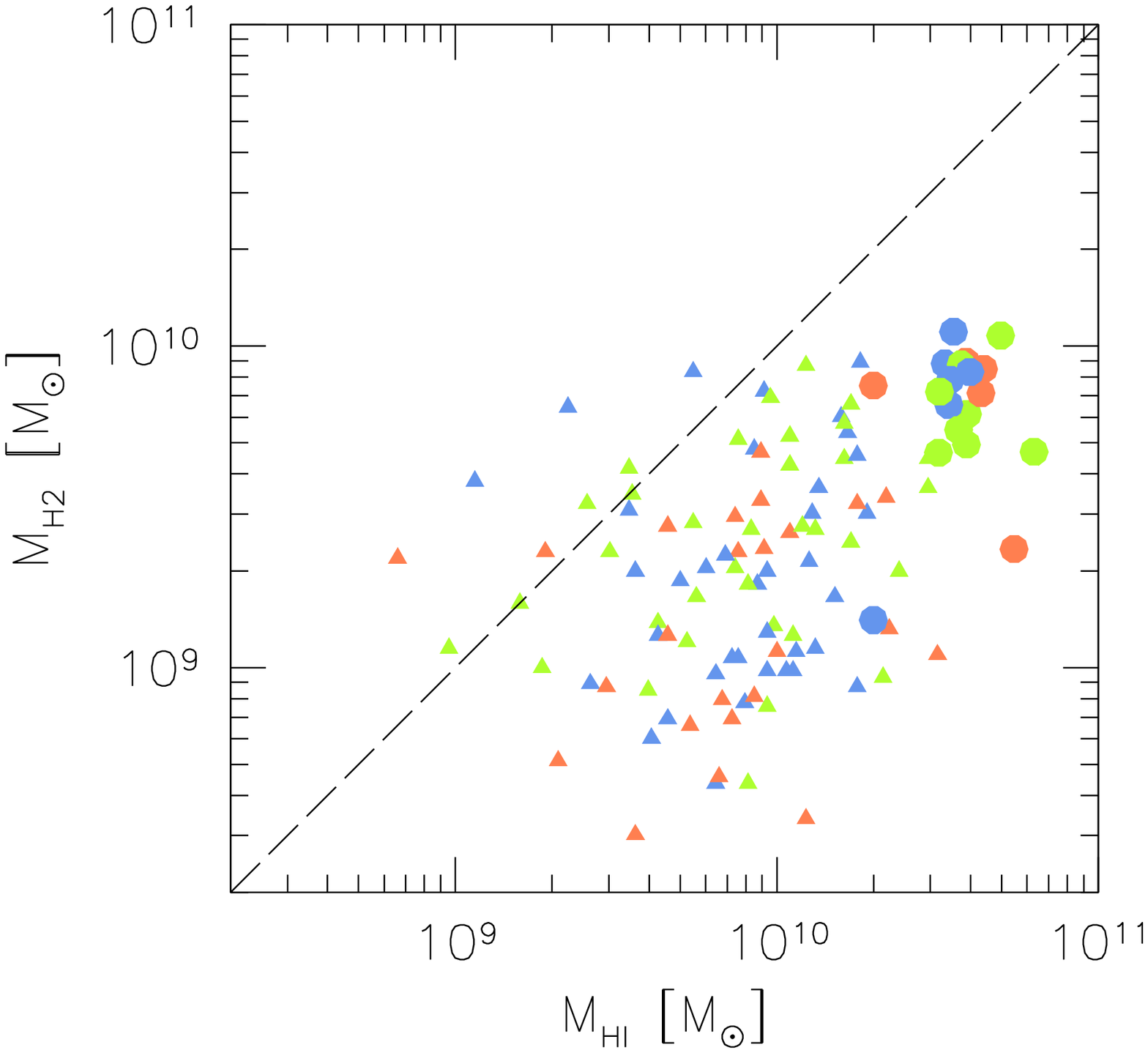,width=0.45\linewidth,clip=}\epsfig{file=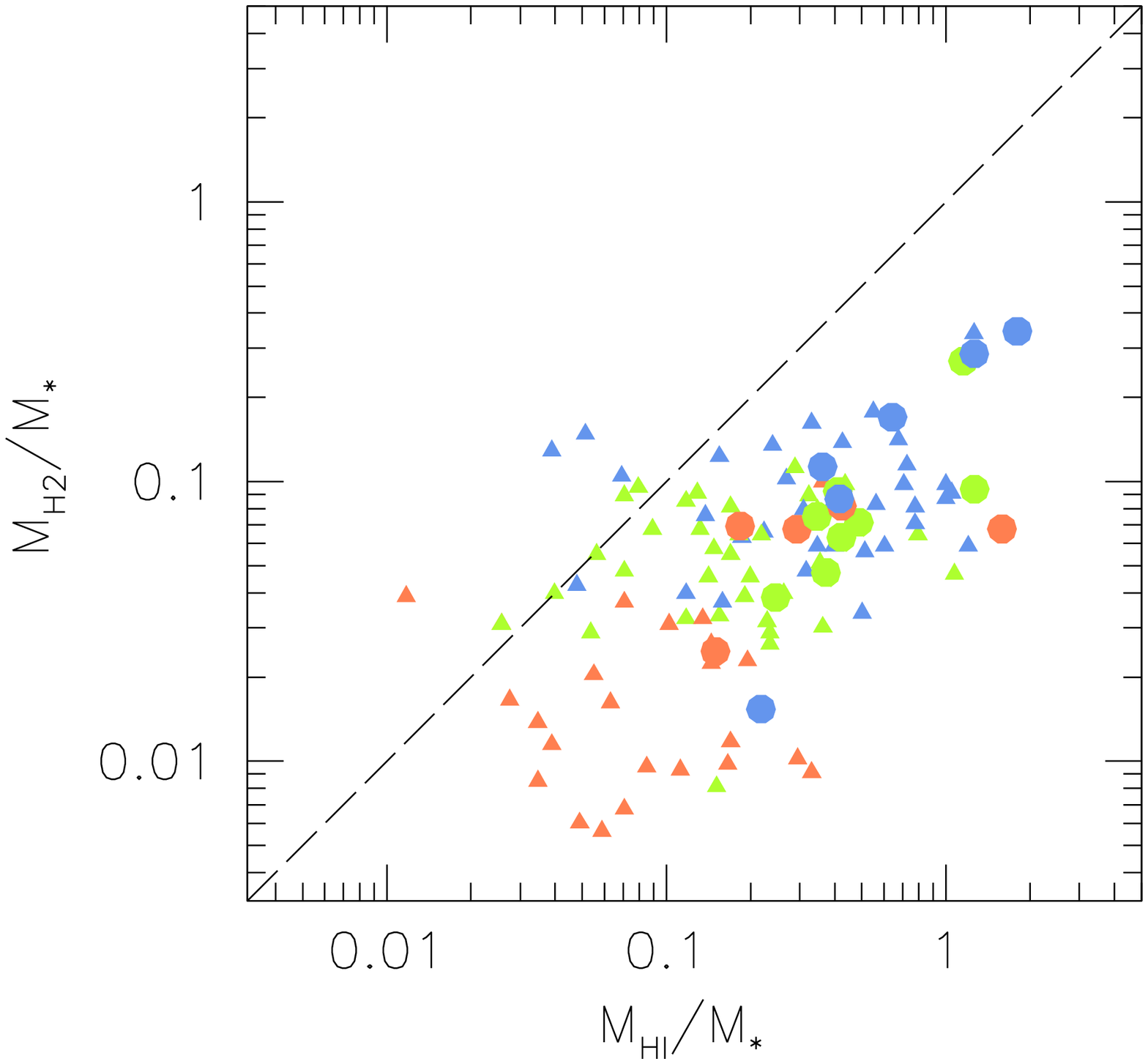,width=0.45\linewidth,clip=}
\vspace{-0.8cm}
\caption{(Left Panel) Molecular gas masses ($M_{\rm H2}$) of the H{\sc i} monsters and the COLD GASS sample are plotted as a function of atomic gas mass ($M_{\rm HI}$). Only the galaxies detected in both H{\sc i} and CO are shown. Symbols are same as Figure~\ref{fig_h2pro}, and their colour represents the $u-r$ colour of the stellar galaxy: orange ($u-r>2.5$), green ($2.1\leq u-r \leq 2.5$), and light-blue ($u-r<2.1$).  The long-dashed line corresponds to the equal mass. 
(Right Panel) A comparison of atomic and molecular gas mass, normalized by stellar mass.    This normalization by stellar mass brings out the systematic difference among galaxies with different star formation history, as represented by the $u-r$ colour.
\label{fig_mhimh2}}
\end{figure*}

In the previous section, we showed that the H{\sc i} monsters are also among the most molecular gas massive systems for their given stellar mass and colour.  This is also clearly seen on the left panel of Figure~\ref{fig_mhimh2}, where H{\sc i} and H$_{2}$ masses of sample galaxies are compared.  Molecular gas mass generally increases with increasing atomic gas mass although the correlation is weak and has a large scatter, as already reported for the COLD GASS sample by \citet{as1}. Galaxies within a similar range of H{\sc i} show at least an order of magnitude spread in their H$_2$ content.  Few galaxies have molecular gas mass exceeding their atomic gas mass, and $M_{\rm HI}$ is almost always larger than $M_{\rm H2}$ by a factor of 2-3 on average. The broad correlation between atomic and molecular gas mass (albeit with a significant scatter), as well as the general trend of a larger atomic gas fraction \citep[i.e., $<M_{\rm H2}/M_{\rm HI}>\sim 0.30$, ][]{as1}, support the widely accepted notion that H{\sc i} is the main reservoir that fuels the formation of dense and cold molecular hydrogen clouds \citep[e.g.][]{s99}.  Our new data extends this trend to galaxies with the highest H{\sc i} masses. 

It is intriguing to find that the scatter among the \hi\ monsters appears to be smaller than those of the COLD GASS galaxies in Figure~\ref{fig_mhimh2}.  This could be simply a result of small number statistics, but a plausible physical explanation emerges when the gas contents are compared after normalizing by stellar mass, as shown on the right panel of Figure~\ref{fig_mhimh2}.  The relation between atomic and molecular gas mass tightens when normalized by stellar mass, and a nearly linear trend emerges.  This result indicates that a large atomic gas reservoir is a clear requirement to produce a large molecular disk in most cases.   The \hi\ monsters appear to show a tighter correlation than the COLD GASS sample in this comparison as well.  The scatter noticeably increases below $M_{\rm HI}/M_* \lesssim 0.2$ among the COLD GASS sample, and colour-coding the galaxies by their $u-r$ colour clearly shows that this increased scatter is driven by a systematic change in the average molecular gas fraction among galaxies with different optical colour.   

\begin{figure*}
\centering
\epsfig{file=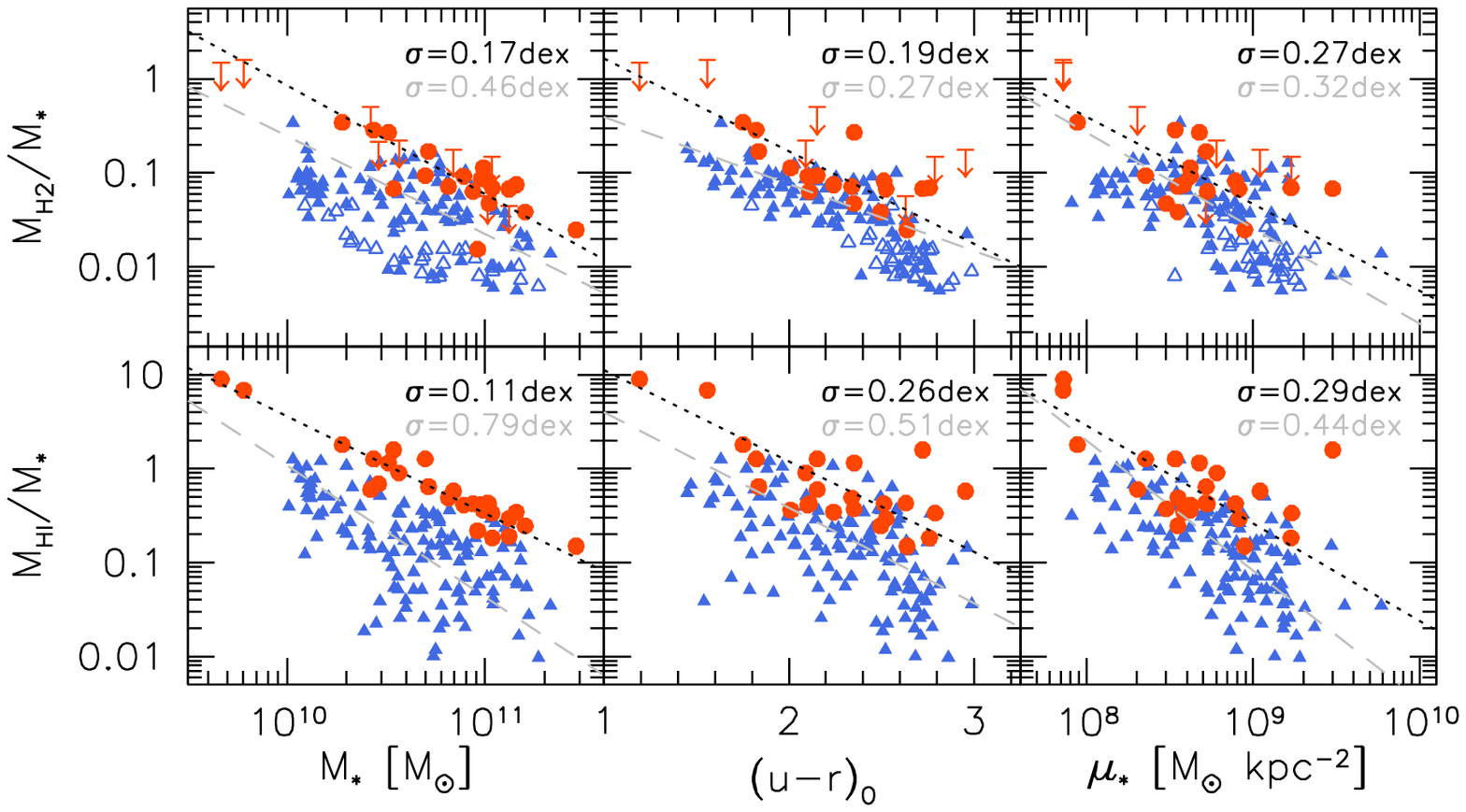,width=1.0\linewidth,clip=}
\vspace{-0.7cm}
 \caption{Molecular and atomic gas mass normalized by stellar mass ($M_{\rm H2}/M_{\ast}$-upper row and $M_{\rm HI}/M_{\ast}$-lower row) as a function of stellar mass $M_{*}$ (left), $u$ -- $r$ colour corrected for galactic extinction ($u$ -- $r$)$_0$ (middle), and stellar surface mass density $\mu_*$ (right). Red circles and blue triangles represent our H{\sc i} monster sample and COLD GASS galaxies, respectively. Red down-arrows and blue open triangles indicate the upper limits for non detections in CO of our sample and the { reference sample}. The grey dashed line in each panel is a linear bisector fit to both samples. For H$_2$-relation plots, only CO detections have been considered for fitting. The black dotted line is a linear bisector fit to H{\sc i} monsters. Again, only CO detections are included for H$_2$ related fitting. The scatters of CO detected H{\sc i} monsters with respect to the fitted lines can be found in the top-right corner of each panel in the same colour as the line. The two extreme outliers among the H{\sc i} monsters that are sticking out in the top-left of every plot of are AGC192542 and AGC262058.}
\label{fig_optgfr}
\end{figure*}

A systematic dependence of molecular gas mass fraction ($M_{\rm H2}/M_*$) on the $u-r$ optical colour is shown in the upper middle panel of Figure~\ref{fig_optgfr}.  A similar trend between molecular gas mass fraction and $NUV-r$ colour was also previously reported by \citet[][see their Fig.~5]{as1}.  These optical colours are usually interpreted as an indicator of different star formation history or specific star formation rate, and star formation activity associated with the molecular gas likely plays a role in the observed trend. The efficiency of converting atomic gas into molecular form is also thought to be regulated by various galactic properties \citep[e.g., mid-plane hydrostatic pressure, metallicity, dust-to-gas ratio, galaxy rotation, radiation field, etc. -- see ][]{br06} as well as accretion/quenching events \citep[galaxy interactions, mergers, stellar and AGN feedback, etc. -- see][and references therein]{pst13}. Regardless of the true mechanism behind this trend, the smaller dispersion in $M_{\rm H2}/M_{\rm HI}$ among the \hi\ monsters appears to be real.
%primarily driven by the fact that the majority of \hi\ monsters are star forming galaxies with blue optical colour.  
Possible dependence on star formation activities and stellar mass density are discussed further below.

\section{Discussion}
\label{discussion}
In this section, we discuss how the molecular gas and total gas masses of the H{\sc i} monsters correlate with various stellar properties (stellar mass, stellar surface mass density, $u$ -- $r$ colour) and star formation activities.  Being most gas-massive galaxies (and gas-richest in most cases) in the Local Universe, these \hi\ monsters offer the most stringent test to various physical correlations identified between gas and stellar properties and could in turn offer valuable insights on physical processes that govern these trends.  We also touch on the implications on theoretical studies of cold gas content in galaxies.

\subsection{Molecular gas and optical properties}
\subsubsection{Stellar mass and colour}
\label{Stellar mass and colour}

Among the H{\sc i} monsters, $M_{\rm H2}/M_*$ varies significantly over two orders of magnitude, ranging from $\lesssim$0.015 (AGC009515) to 0.345 (AGC222077). The median value of $M_{\rm H2}$/M$_{\ast}$ for the CO detected H{\sc i} monsters is 0.075, which is similar to the average $M_{\rm H2}/M_*$ of the COLD GASS galaxies, $0.066 \pm 0.039$ \citep{as1}.  
Molecular gas masses of the \hi\ monsters and the comparison sample of COLD GASS galaxies normalized by stellar mass ($M_{\rm H2}/M_*$) are shown along the upper row of Figure~\ref{fig_optgfr}. 
H{\sc i} monsters are found at the upper end of the normalized \htwo\  mass distribution for a given stellar mass and $u-r$ colour, which is also true for the normalized H{\sc i} mass (bottom panels). This confirms the conclusion from \S\ref{relation_h1_h2} that our H{\sc i}-massive galaxies also tend to have a large \htwo\  gas reservoir.

After analyzing the molecular and atomic gas contents of the COLD GASS sample, \citet{as1} have reported that the mean molecular gas fraction $M_{\rm H2}/M_*$ does not strongly depend on stellar mass, stellar surface mass density, or concentration index, while it has a strong dependence on the $NUV-r$ colour. Our analysis of the combined \hi\ monsters  and COLD GASS sample shown in Figure~\ref{fig_optgfr} confirms the earlier conclusions by \citet{as1}, including a correlation between $M_{\rm H2}/M_*$ and $u-r$ colour.  This correlation indicates that molecular gas content is indeed closely tied to galaxy stellar properties and that a large dispersion in molecular gas fraction as a function of $M_*$ and $\mu_*$ reflects  a wide spectrum of star formation history associated with individual galaxies, as indicated by their optical colour  \citep[e.g.][]{b4}.  This connection between the gas properties and star formation activity is examined further below in section \S~\ref{sec:sfr}.

The scatter in the broad trends seen in all six panels of Figure~\ref{fig_optgfr} is systematically smaller for the \hi\ monsters compared with the COLD GASS sample.  The tight correlation between $M_{\rm HI}/M_*$ and $M_*$ seen in the bottom left panel is almost certainly due to the sample selection, as the \hi\ monsters represent the most \hi-massive galaxies with a narrow range of $M_{\rm HI}$ (see Fig.~\ref{fig_hipro}).  A broad correlation between $M_{\rm HI}$ and $M_{\rm H2}$ (Fig.~\ref{fig_mhimh2}) offers a natural explanation for the systematic displacement of the \hi\ monsters on the higher \htwo\ mass fraction side of the mean trends in the upper panels, while a large scatter in the $M_{\rm HI}$-$M_{\rm H2}$ correlation translates to a larger dispersion in the \htwo\  mass fraction trends.  Given the tighter correlation between $M_{\rm HI}/M_*$ and $M_{\rm H2}/M_*$ (right panel in Fig.~\ref{fig_mhimh2}), a better overlap between the \hi\ monsters and the COLD GASS sample is expected in the correlation between \htwo\ mass fraction and $u-r$ colour.  

Indeed the two samples overlap better, but a small systematic offset persists.  As discussed in some length by \citet{as1}, a comparison of CO data from different telescopes can be problematic with possible systematic differences in calibration and aperture correction.  The relative calibration between our data and the IRAM 30~m telescope seems quite good, at least for compact sources, as demonstrated by the comparison of the measured CO line integrals for a sample of ultra-luminous infrared galaxies as shown in Figure~5 of \citet{c09}.  Stellar disks of some of the \hi\ monsters are slightly larger than the 50$''$ beam of the RSR (see Fig.~\ref{fig_rsrsp}), and the resulting \htwo\ masses may be a slight {\em under-estimate} in some cases.  In contrast, the COLD GASS sample galaxies were observed with a beam only 22$''$ in size, requiring an empirical aperture correction in nearly all cases \citep{as1}.  The COLD GASS sample and \hi\ monsters overlap better on the redder end ($u-r>2$), and the apparent difference among late type galaxies may reflect either a true systematic shift in balance among atomic, molecular, and stellar mass density (see \S~\ref{sec:sfr}) or a systematic error in the aperture correction by Saintonge et al. among these late type galaxies.

\subsubsection{Stellar and gas surface mass density}
\label{stellar mass density and conc}

\begin{figure}
\centering
\vspace{-0.5cm}
\epsfig{file=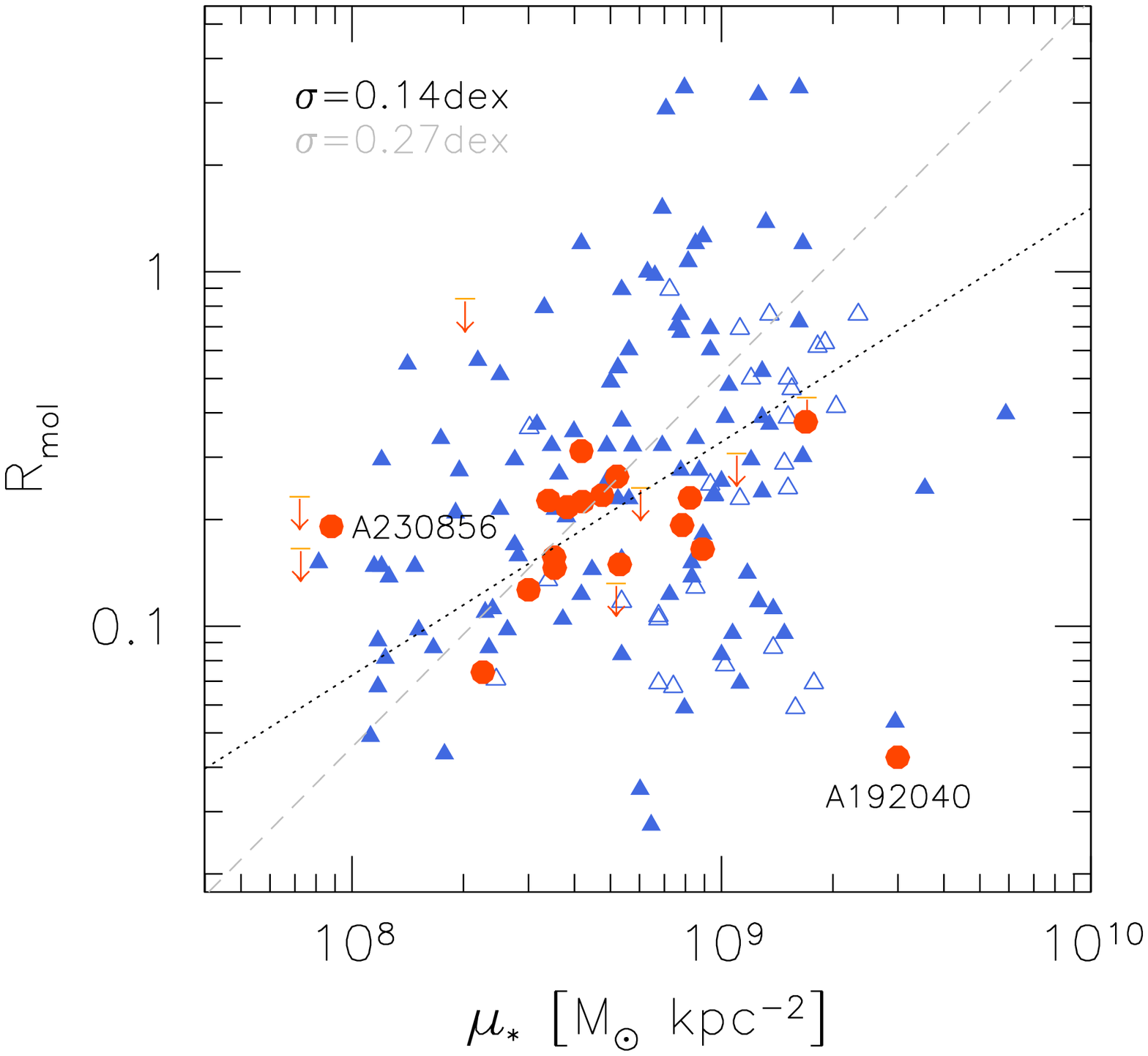,width=1.0\linewidth,clip=} 
\vspace{-1.2cm}
\caption{Stellar mass surface density $\mu_*$ vs. $R_{mol}=M_{\rm H2}/M_{\rm HI}$. Symbols are same as Figure~\ref{fig_optgfr}. The dashed line is a linear bisector fit to all galaxies in both samples, and the scatter about this broad trend is large, $\sigma=0.39$. The dotted line is a linear bisector fit to CO detections of H{\sc i} monsters excluding two extreme outliers (labeled in the figure), and the scatter among the \hi\ monsters excluding the two outliers is 0.14 dex.}
\label{fig_stdgsd}
\end{figure}

\citet{as1} found little dependence between molecular gas fraction ($M_{\rm H2}$/$M_\ast$) and stellar surface mass density ($\mu_\ast$), and this is also seen among the H{\sc i} monsters and the COLD GASS galaxies in Figure~\ref{fig_optgfr}.  This is somewhat surprising since \citet{br06} have found a tight correlation between molecular-to-atomic gas mass ratio $R_{mol} \equiv \mu_{\rm H2}/\mu_{\rm HI}$ and the mid-plane hydrostatic pressure $P_{\rm ext}$, which is controlled by the stellar surface mass density $\mu_*$. In Figure~\ref{fig_stdgsd}, we compare $R_{mol}$ and $\mu_*$ for the H{\sc i} monsters and the COLD GASS galaxies, and again no correlation is recognizable for the full sample, contrary to the expectations from the pressure-driven gas phase transition scenario.

The  H{\sc i} monsters by themselves cluster together with a possible underlying correlation between $R_{mol}$ and $\mu_*$ in Figure~\ref{fig_stdgsd}. Note that the ratio between $\mu_{\rm HI}$ and $\mu_{\rm H2}$ is essentially the molecular-to-atomic gas mass ratio ($=M_{\rm H2}/M_{\rm HI}$) when one characteristic size is adopted for individual galaxies, which is the case here. The grey dashed line is a linear bisector fit to all H{\sc i} and CO detected galaxies in both samples, and the scatters around this line of H{\sc i} monsters and COLD GASS galaxies are 0.27 dex and 0.39 dex, respectively.  The dotted line represents a bisector fit only for the \hi\ monsters excluding one outlier - AGC192040, and the correlation seems to be much stronger with $\sigma$ of only 0.14 dex. 
The outlier, AGC192040 has a compact/red central component with ring-like stellar arms, and its morphology is not typical of a disk galaxy. This system might be the result of recent merging and thus not follow the same pressure-density relation as normal spirals. AGC230856, which appears to be a normal spiral yet with unusually blue colour (the bluest after two least massive galaxies), is also labelled but still included in our statistics. This galaxy is likely to have gone through a distinct evolution from normal spirals and natural to deviate from other gas-massive galaxies in the opposite sense from the AGC192040's case.

\begin{figure*}
\centering
\epsfig{file=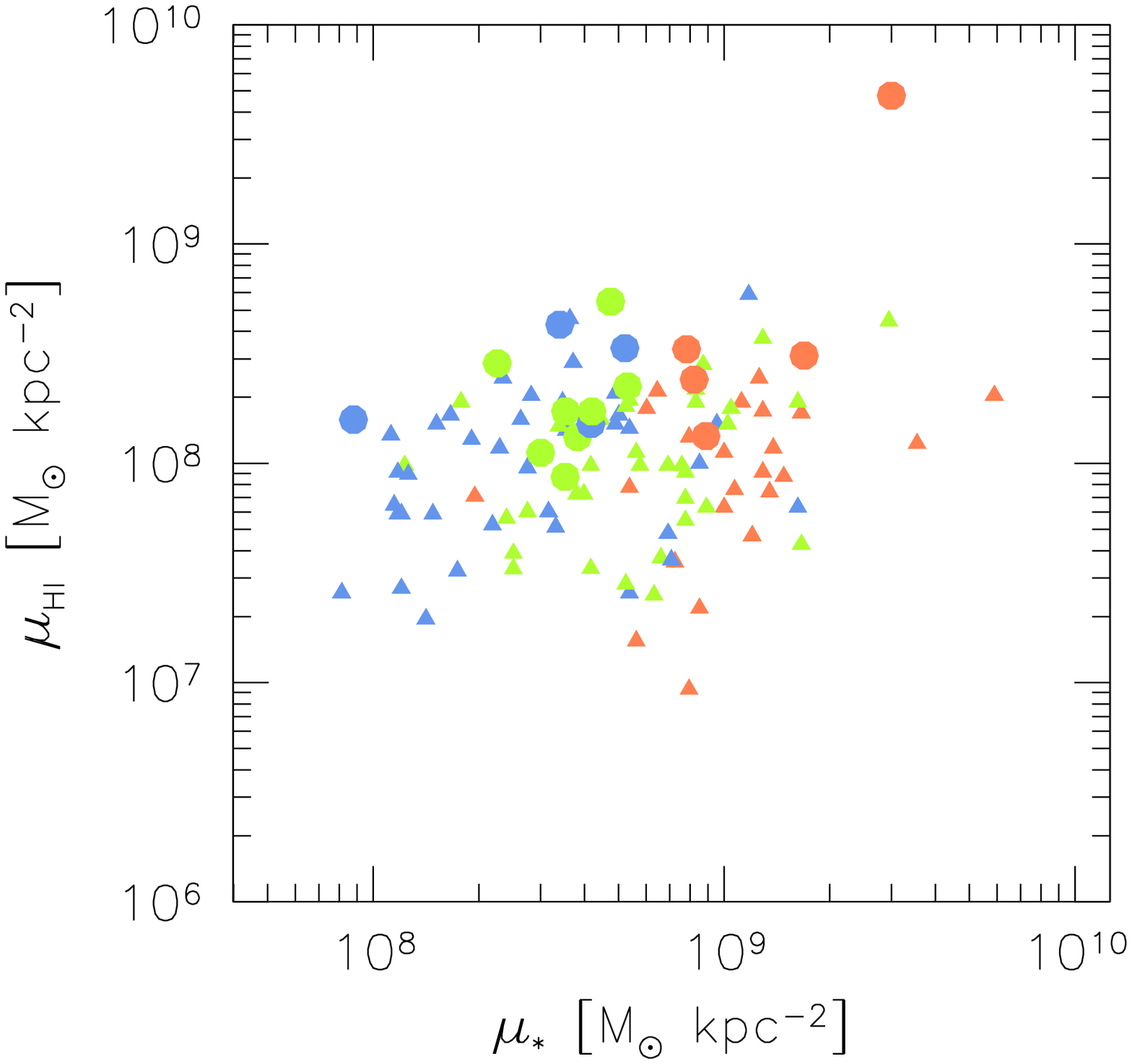,width=0.45\linewidth,clip=}\epsfig{file=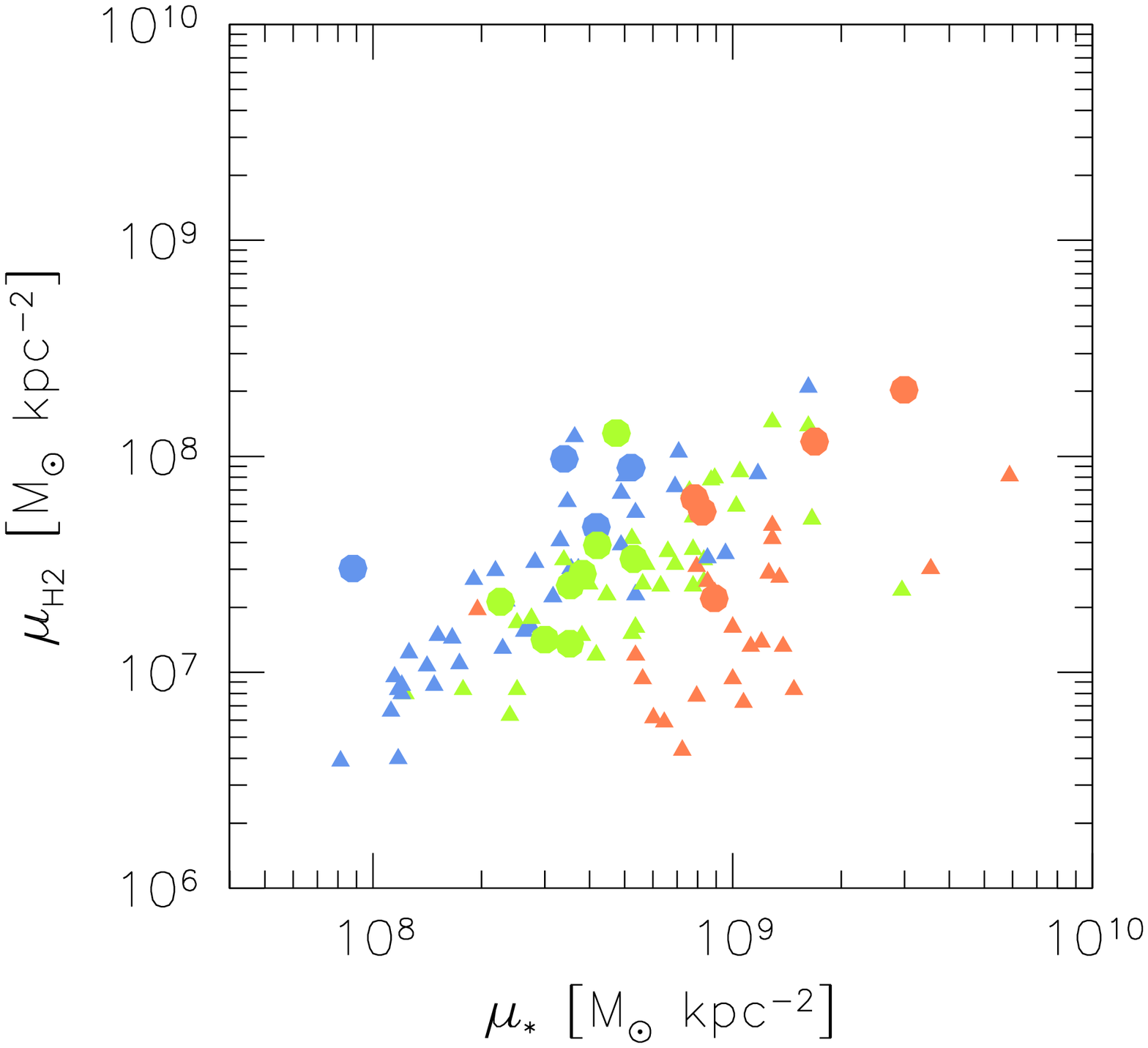,width=0.45\linewidth,clip=} 
\vspace{-0.8cm}
 \caption{Mean atomic gas surface density $\mu_{\rm HI}$ (left) and molecular gas surface density $\mu_{\rm H2}$ (right) of the H{\sc i} monsters and COLD GASS galaxies, which are defined in the same manner as $\mu_*$, following eq.~(1), are shown as a function of stellar mass density $\mu_*$. Symbols and colours are the same as Figure~\ref{fig_mhimh2}. 
 \label{fig_sigmas}}
\end{figure*}

A clue to the lack of any correlation between molecular mass fraction ($M_{\rm H2}/M_*$) and molecular-to-atomic gas mass ratio ($R_{mol}$) with stellar surface mass density $\mu_\ast$ in Figures~\ref{fig_optgfr} \& \ref{fig_stdgsd} emerges when atomic and molecular surface mass density ($\mu_{\rm HI}$ and $\mu_{\rm H2}$ - computed in the same way as $\mu_{\ast}$) are compared with $\mu_\ast$ as shown in Figure~\ref{fig_sigmas}. A comparison of atomic gas and stellar mass density, shown on the left panel, again shows no correlation, suggesting that these two quantities are independent of each other.  Galaxies with different $u-r$ colour separate themselves in stellar mass density $\mu_*$, and the dispersion in atomic gas density $\mu_{\rm HI}$ is equally large across the full range of $\mu_*$.   On the other hand, a comparison of $\mu_{\rm H2}$ with $\mu_\ast$ on the right panel shows a clear correlation, indeed suggesting the correlation between the {\em molecular gas density} and the stellar mass density. This correlation becomes significantly tighter when the galaxies are sorted by their $u-r$ colour, suggesting that galaxy colour is an important secondary parameter underlying this correlation between $\mu_{\rm H2}$ and $\mu_{\ast}$.  Galaxies with blue colour ($u-r < 2.5$) in particular forms a tight trend with $\sigma \lesssim 0.2$ in dex, while redder galaxies also show a similar trend but displaced in $\mu_\ast$ by a factor of $\sim$5. 

The absence of any correlation between $\mu_{\rm HI}$ with $\mu_\ast$ and a tight correlation between $\mu_{\rm H2}$ with $\mu_\ast$ with a strong $u-r$ colour dependence shown in Figure~\ref{fig_sigmas} offers a natural explanation for the lack of any correlation between molecular-to-atomic gas mass ratio ($R_{mol}$) with stellar surface mass density $\mu_\ast$ among the COLD GASS sample in Figure~\ref{fig_stdgsd} and the apparent correlation in the same plot for the \hi\ monsters.  The former is a trivial outcome of $\mu_{\rm HI}$ being independent of $\mu_\ast$, as $R_{mol}=\mu_{\rm H2}/\mu_{\rm HI}$ should also be independent of $\mu_\ast$, regardless of any correlation between $\mu_\ast$ and $\mu_{\rm H2}$.   Given that, the apparent correlation among the \hi\ monsters is somewhat unexpected.  It turns out this correlation arises because the \hi\ monsters have a narrow range of \hi\ masses by the sample definition, similar to the apparent tight trend seen between $M_{\rm HI}/M_*$ and $M_*$ in Figure~\ref{fig_optgfr}.   In other words, the apparent correlation between $R_{mol}$ and $\mu_\ast$ among the \hi\ monsters also reflects the correlation between $\mu_{\rm H2}$ and $\mu_\ast$.  Some of the \hi\ monsters show a slightly elevated level of $\mu_{\rm H2}$ for their given $\mu_*$ and colour (as discussed in \S~\ref{Stellar mass and colour}), but this enhancement is not enough to disrupt the observed correlation.

The surface mass densities $\mu_{\rm HI}$, $\mu_{\rm H2}$ and $\mu_\ast$ discussed here are {\em global} averages, normalized by the characteristic size of their stellar disk, and this is an important difference from the quantities considered by \citet{br06} in explaining the physical connection between {\em local} molecular-to-atomic gas mass ratio ($R_{mol}$) with {\em local} stellar surface mass density (or pressure).  The observed correlation between $\mu_{\rm H2}$ with $\mu_\ast$ in Figure~\ref{fig_sigmas} suggests that the scaling relation between molecular gas and stellar disk holds well even at global scales, although there is also a clear systematic dependence on the colour (and thus stellar population and possibly spatial distribution).  The absence of correlation between {\em globally averaged} $\mu_{\rm HI}$ with $\mu_\ast$ indicates the disconnect in spatial distribution between atomic gas and stellar disk \citep[i.e. $R_{\rm HI}>R_{\rm H2}$, see][]{walter08}, as well as between total \hi\ and stellar mass, as discussed earlier (in \S~\ref{monsters}).  %Future spatially resolved imaging of \hi\ distribution should be able to differentiate these two scenarios.  Regardless, 
The breakdown of the correlation between gas and stellar surface mass densities on global scales has an important consequence in the modeling of the gas content and star formation evolution in galaxies as we discuss further below.

\begin{figure*}
\centering
\epsfig{file=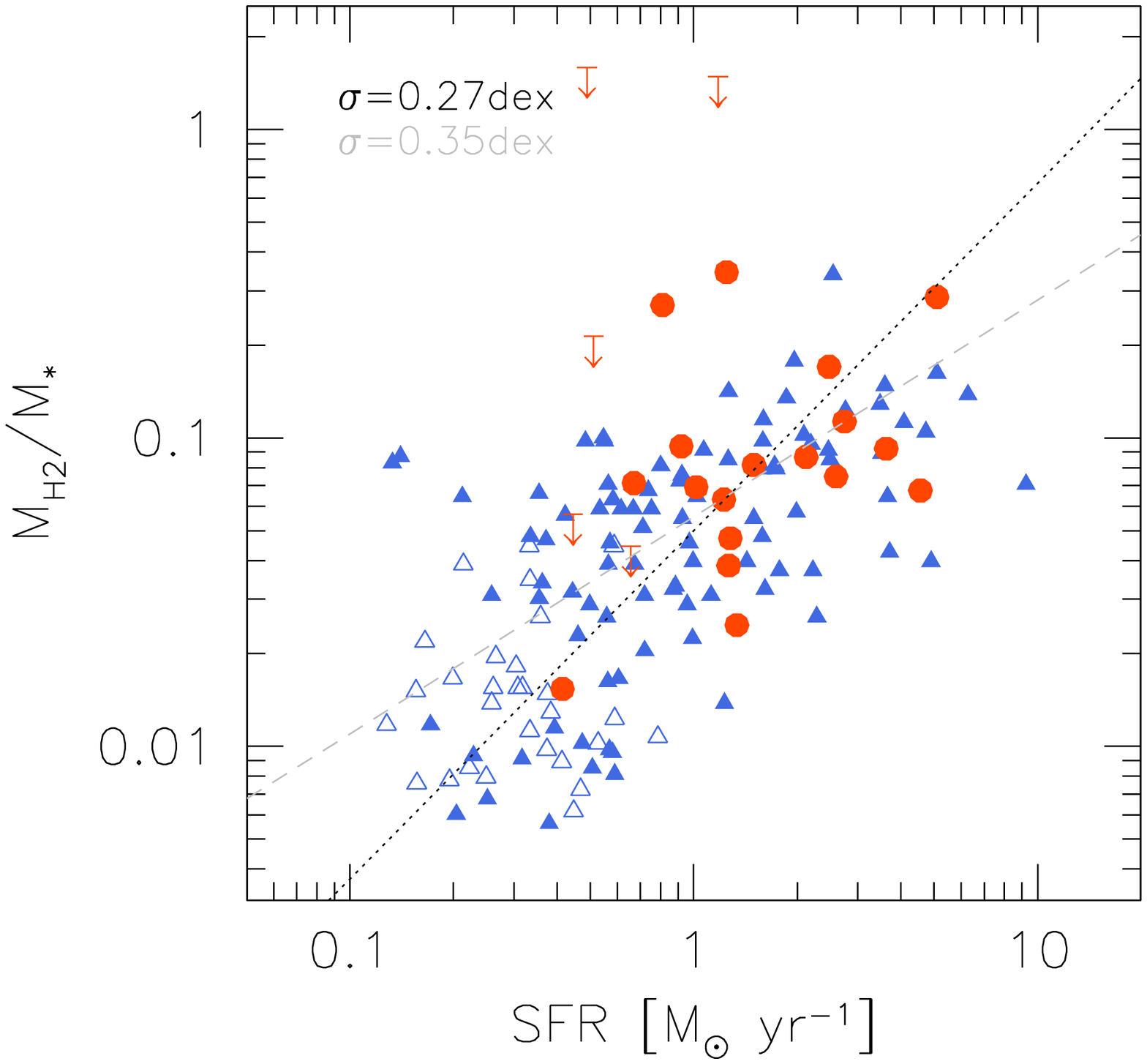,width=0.5\linewidth,clip=}\epsfig{file=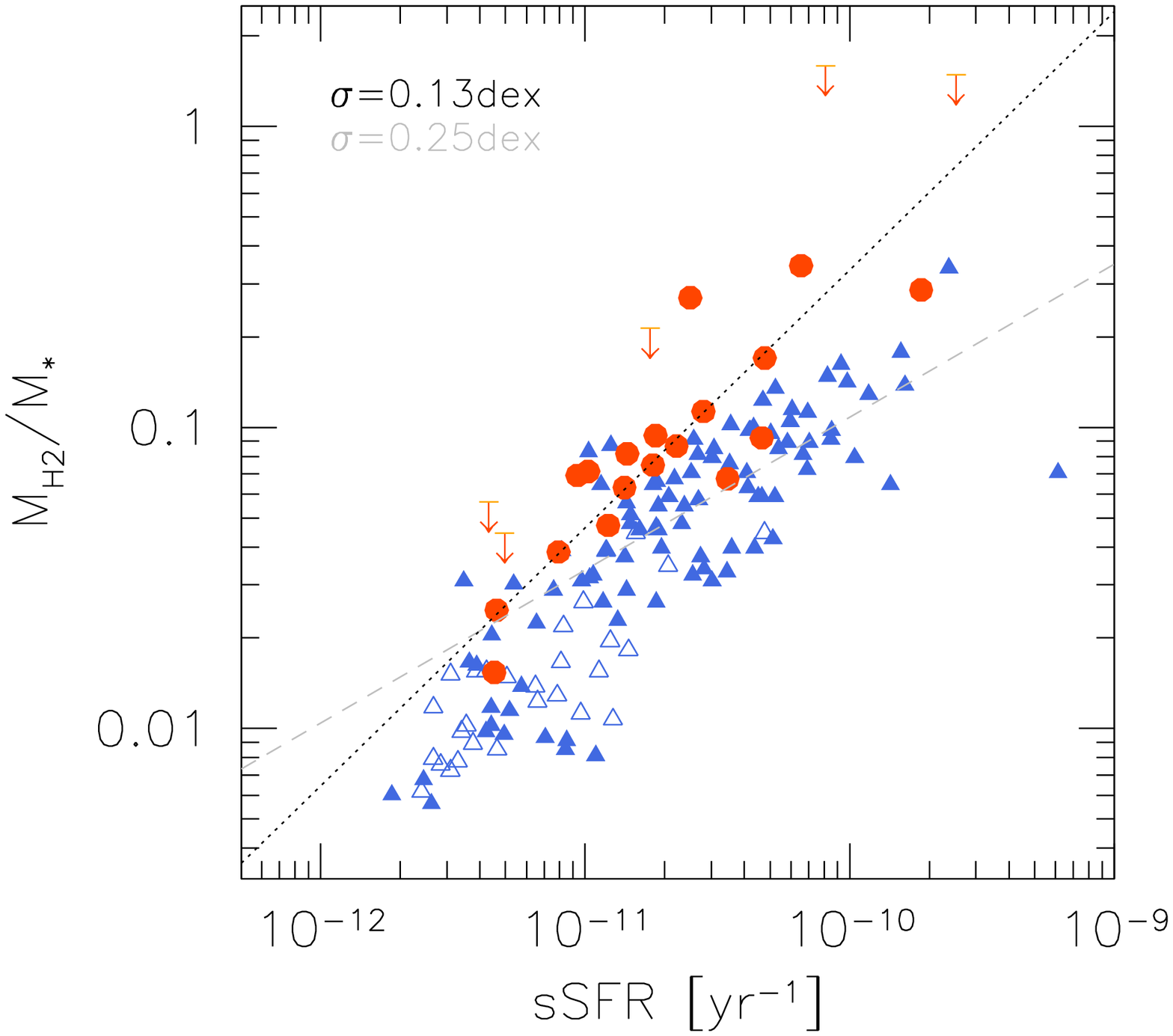,width=0.5\linewidth,clip=} 
\vspace{-1.2cm}
 \caption{Molecular gas fraction ($M_{\rm H2}/M_*$) as a function of star formation rate (SFR) and specific star formation rate (SSFR=SFR$/M_*$). Symbols and lines are same as Figure~\ref{fig_optgfr}. \label{fig_sfpro}}
\end{figure*}

\subsection{Gas content and star formation history}
\subsubsection{Star formation rate and molecular gas content}
\label{sec:sfr}

As noted previously by \citet{as1},  the tightest correlation seen among the comparisons of different gas and stellar properties of the \hi\ monsters and the COLD GASS sample is also the tight relation between molecular gas fraction $M_{\rm H2}/M_*$ and $u$ -- $r$ colour (upper middle panel in Fig.~\ref{fig_optgfr}). In the previous section, we found a clear correlation between molecular gas and stellar surface density (right panel of Fig.~\ref{fig_sigmas}) while no correlation is found between $\mu_{\rm HI}$ and $\mu_*$ (left panel), suggesting a physical disconnect between spatial distribution (and/or relevant time scales) of atomic gas and stellar component.  Taken together, these observations suggest a close physical link between star formation activity primarily with molecular gas content of galaxies on global scales.  Earlier CO surveys have reported a tight correlation between CO and far-$IR$ luminosity among normal and starburst galaxies, and this is interpreted similarly as a scaling relation between the amount of molecular gas and the luminosity from young stars forming within \citep[see review by][]{young91}.

To explore this idea further, we have compiled optically derived star formation rate (SFR) estimates from the MPA-JHU DR7 database\footnote{http://www.mpa-garching.mpg.de/SDSS/DR7/sfrs}, which uses the SDSS spectra and photometry and the method described by \citet{brinchmann}.  We have also derived $IR$-based estimate of SFR using the WISE 22 \micron\ photometry (see \S~\ref{quantities}).  These two different estimates show a good agreement within the SFR range of 0.1 and 10 $M_\odot$ yr$^{-1}$ represented by the \hi\ monsters and the COLD GASS sample, although the dispersion in the correlation is significant ($\sigma \sim 0.3$ in dex).  A comparison of molecular gas fraction versus star formation rate and specific star formation rate (sSFR) shows a much tighter correlation with the $IR$-derived quantities (e.g., $\sigma=0.25$, compared with $\sigma=0.38$ for the SDSS).  Therefore, we limit our discussion to the $IR$-derived star formation quantities from this point on.

Molecular gas fraction and star formation rate SFR track each other broadly, as shown on the left panel of Figure~\ref{fig_sfpro}. Galaxies with a higher SFR tend to have a larger fractional molecular gas content, and the \hi\ monsters are nearly indistinguishable from the comparison COLD GASS sample. The scatter among individual galaxies is still large for both samples ($\sigma=0.27$ for the \hi\ monsters, $\sigma=0.35$ for the whole sample), but this correlation is clearly seen spanning all galaxies of different types and colours (red galaxies to the bottom left and blue galaxies to the top right).  

Specific star formation rate is even more tightly correlated with the molecular gas fraction (see the right panel), having the smallest value of $\sigma=$ 0.13 dex for the \hi\ monsters and $\sigma=$ 0.25 dex for the entire sample.  Most H{\sc i} monsters follow the same relation as the COLD GASS galaxies, although the systematic offset from the { reference sample} is visible again (see below).  %As already discussed in \S~\ref{Stellar mass and colour}, this offset may reflect the systematically larger overall gas content among the \hi\ monsters by the sample selection. 
A tighter correlation between $M_{\rm H2}/M_*$ and sSFR, rather than with SFR, is somewhat unexpected, but this appears to be a different realization of the linear scaling relationship between the amount of molecular gas and the amount of young stars that are forming within them.  Since specific star formation rate is by definition SFR normalized by $M_*$, the correlation seen on the right panel of Figure~\ref{fig_sfpro} is the correlation between $M_{\rm H2}$ and $SFR$, both normalized by $M_*$ to remove the distance dependence. This correlation shows the smallest dispersion among all comparisons of physical quantities we have examined for the \hi\ monsters and the COLD GASS sample, and this is likely the most fundamental relation that threads all types of galaxies represented in this study.

The small systematic offset seen between the \hi\ monsters and the { reference sample} on the right panel of Figure~\ref{fig_sfpro} is surprising, given the tightness of the correlation ($\sigma=0.13$).  Besides the \hi\ monsters, several other COLD GASS galaxies are also offset from the main trend, in the manner that can be interpreted as having a SFR 2-3 times {\em smaller} for their given molecular gas mass.  As discussed in \S~\ref{Stellar mass and colour}, this offset may largely reflect the sample selection of  galaxies with the largest cold gas content.  However, we already established no physical link between \hi\ content and SFR (\S~\ref{stellar mass density and conc}), and a {\em lower SFR for given molecular gas mass among the H{\small I} monsters} requires a different explanation.  %An explanation might be found in different time scales involved.  The tight correlation between $M_{\rm H2}$ and SFR discussed in the previous paragraph would require that the time scale for young stars is sufficiently short ($\le 10^7$ yr) compared with the molecular cloud lifetime ($\sim10^8$ yr) so that SFR tracks $M_{\rm H2}$ nearly instantaneously.  
Cold gas content and consumption is only a quasi-static balance among gas accretion from the halo to the disk, star formation, recycling of the star-gas cycle, and feedback \citep[e.g., ][]{bau10}.  Small changes in the physical condition may swing this balance between $M_{H2}$ and $SFR$, for example, as seen among the \hi\ monsters, or the very condition that has led to the large \hi\ reservoir may simultaneously lead to the shift in the balance as observed.  These are only conjectures at the moment, but this issue may deserve a further attention in future studies.

\begin{figure*}
\centering
\epsfig{file=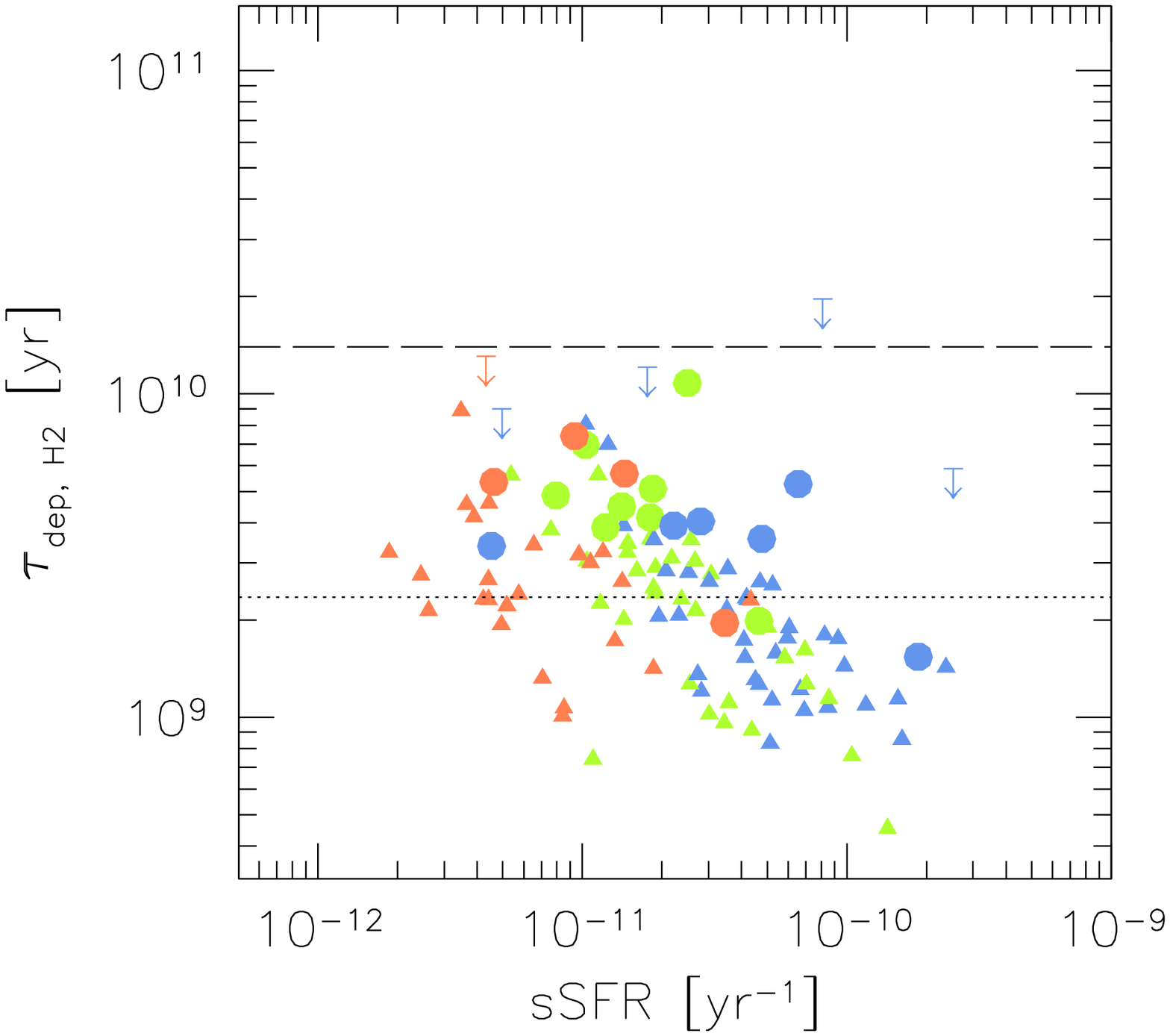,width=0.45 \linewidth,clip=}\epsfig{file=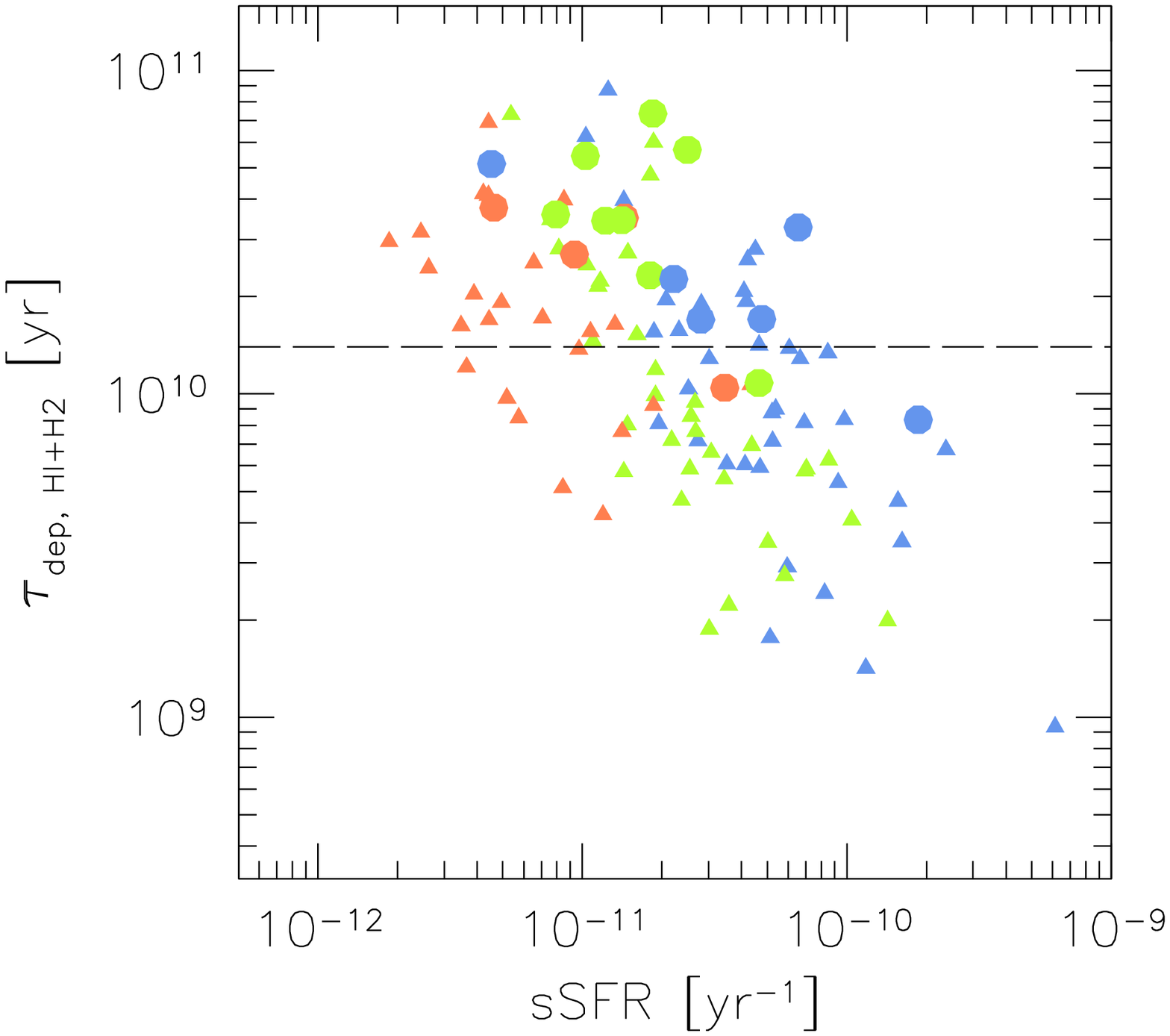,width=0.45 \linewidth,clip=} 
\vspace{-0.8cm}
 \caption{Molecular gas depletion timescale $\tau_{dep,\rm H2}$= $M_{\rm H2}$/SFR (left panel) and total cold gas depletion timescale  $\tau_{dep,\rm HI+H2}$ = ($M_{\rm HI}+M_{\rm H2}$)/SFR (right panel) as a function of specific star formation rate. Symbols and colours are the same as Figure~\ref{fig_mhimh2}.  Long dashed line represents the Hubble time of 14 Gyr.  The dotted line on the left panel corresponds to the constant molecular gas depletion time in nearby disk galaxies by \citet{bigiel11}.}
 \label{fig_ssftim}
\end{figure*}

\subsubsection{Total and molecular gas consumption/depletion time}
\label{sec:tausfr}

One of the main motivations for conducting this study was testing whether a larger neutral atomic gas content (``reservoir") automatically translates to a larger molecular gas mass and a higher SFR, as commonly formulated in cosmological simulations of galaxy growth and cold gas content evolution \citep[e.g., ][]{obr09a,bau10,fu10,lagos11}.  Our new CO survey of \hi\ monsters confirms that galaxies with the largest \hi\ content in the Local Universe also tend to have the top end of the range of molecular gas content among the galaxies of similar stellar mass and colour (see \S~\ref{relation_h1_h2}).  However, H{\sc i} monsters are {\em not} more efficient in forming stars for their abundant molecular gas mass.  In fact, they may be systematically {\em less} efficient in turning their molecular gas into stars, as discussed in the previous section (\S~\ref{sec:sfr}).  This in turn means that the time scale for consuming their entire molecular gas reservoir at the current star formation rate should be longer for the H{\sc i} monsters.  

As seen on the left panel in Figure~\ref{fig_ssftim}, molecular gas depletion time ($\tau_{dep,\rm H2}\equiv M_{\rm H2}/$SFR) is broadly correlated with specific star formation rate with a large scatter.  At a given specific star formation rate, H{\sc i} monsters have a systematically larger value of $\tau_{dep,\rm H2}$ than the { reference sample} COLD GASS galaxies.  The median values of the molecular gas depletion time are 2.0~Gyr and 4.3~Gyr for the CO detected COLD GASS galaxies and H{\sc i} monsters, respectively.  The ``constant" molecular gas depletion time of 2.35 Gyr in nearby spiral galaxies derived by \citet{bigiel11} (dotted line) runs approximately through the middle of the data points.  While the Bigiel et al. report an apparent scatter of $\sim 0.2$ in dex (see their Figure 2) for their spiral galaxies, the spread in $\tau_{dep,\rm H2}$ for the COLD GASS and \hi\ monsters is several times larger than this scatter in the mean relation and has a systematic dependence on sSFR.  A short molecular gas depletion time (compared with the Hubble time, long dashed line), especially for the COLD GASS sample, {\em requires} a continuous replenishment of cold gas from the surrounding halo if their current SFR were to be sustained \citep[e.g., ][]{putman06}.   

We also find that the addition of $M_{\rm HI}$ greatly extends the gas depletion time to close to or exceeding the Hubble time, and the gas supply that can fuel the current level of star formation is readily found in the \hi\ phase in many cases.  Neutral hydrogen in and around galaxies is widely viewed as the reservoir for supplying gas to mostly molecular gas disk and the associated star formation activity  \citep[e.g.,][]{bau10}.  The {\em total} cold gas depletion time ($\tau_{dep,\rm HI+H2}\equiv M_{\rm HI+H2}/$SFR, right panel in Figure~\ref{fig_ssftim}) suggests that nearly all \hi\ monsters and a significant fraction of the \hi\ detected COLD GASS galaxies can sustain their current level of star formation for the Hubble time or longer.   Numerical studies such as by \citet{b13} suggest that the current gas accretion rate onto DM halo is $\le10\%$ of the peak rate during the first 3 Gyr following the Big Bang.  {\em Our analysis shows that the diminished inflow from the $\ge100$ kpc scale halo at the present epoch is not a critical factor in sustaining the current star formation activity in many of these galaxies, and understanding the processes that lead to and regulating the present H{\small I} content may be a more critical step in achieving a better understanding of the evolution of present day galaxies.}  The total cold gas depletion time $\tau_{dep,\rm HI+H2}$ for bluer, late type galaxies with sSFR $\ge 10^{-10.5}$ yr$^{-1}$ is mostly shorter than the Hubble time,  however, and the  most actively star forming galaxies in the Local Universe still require a significant inflow of fresh gas from their halo in order to sustain their current SFR \citep[also true for the Milky Way Galaxy, see ][]{putman06}.  

\begin{figure*}
\epsfig{file=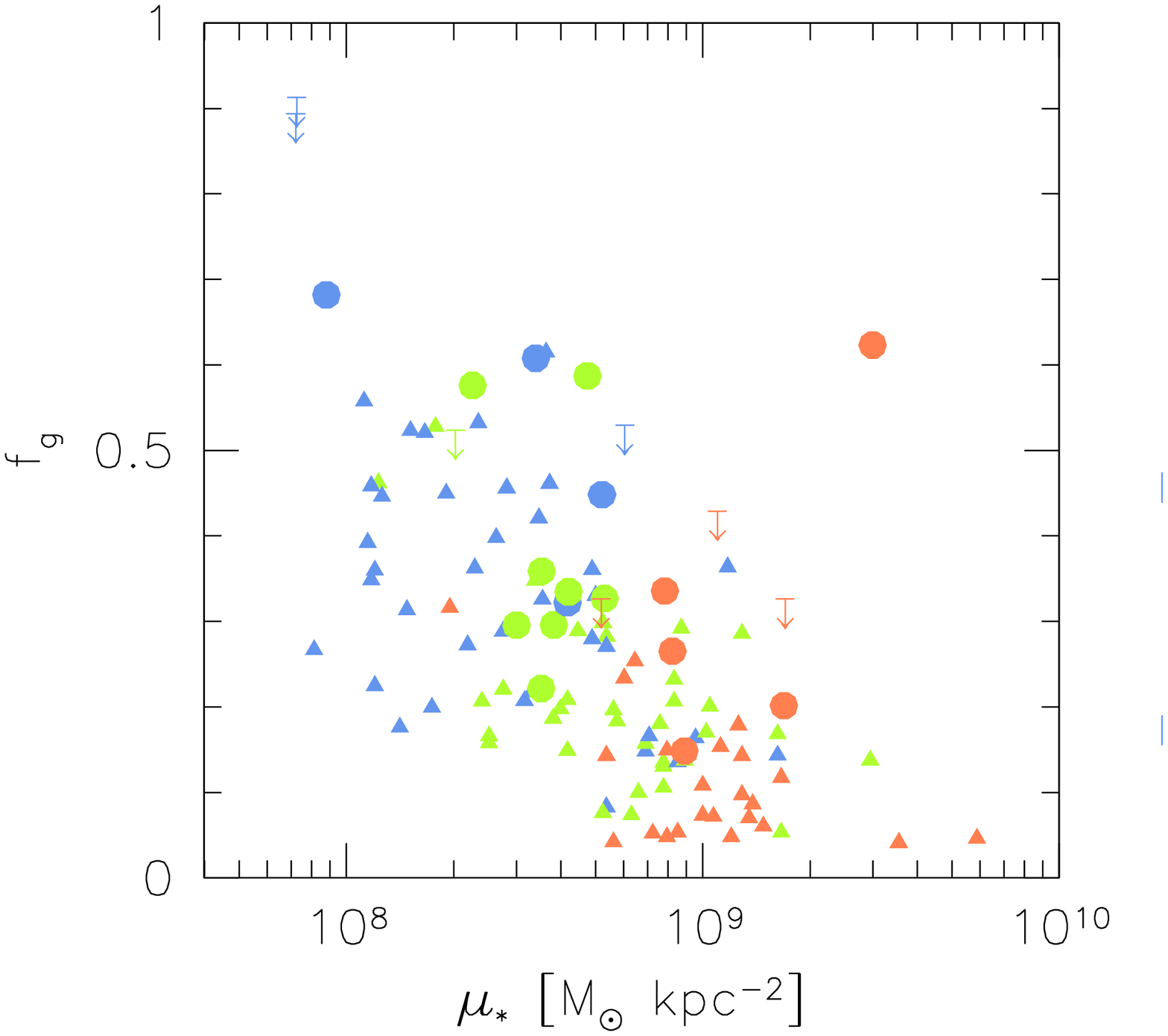,width=0.34\linewidth,clip=}
\hspace{-0.5cm}\epsfig{file=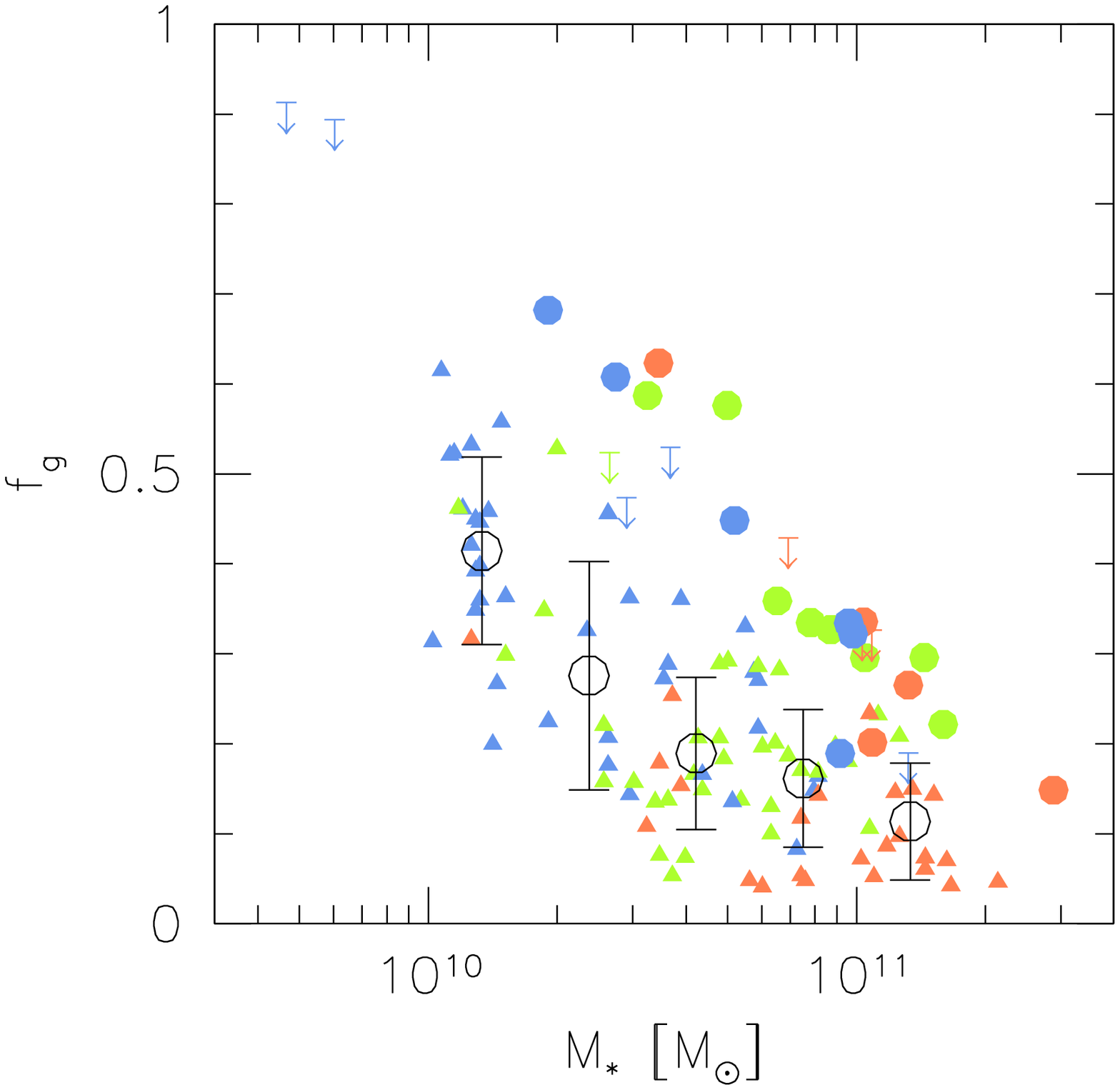,width=0.34\linewidth,clip=} 
\hspace{-0.5cm}\epsfig{file=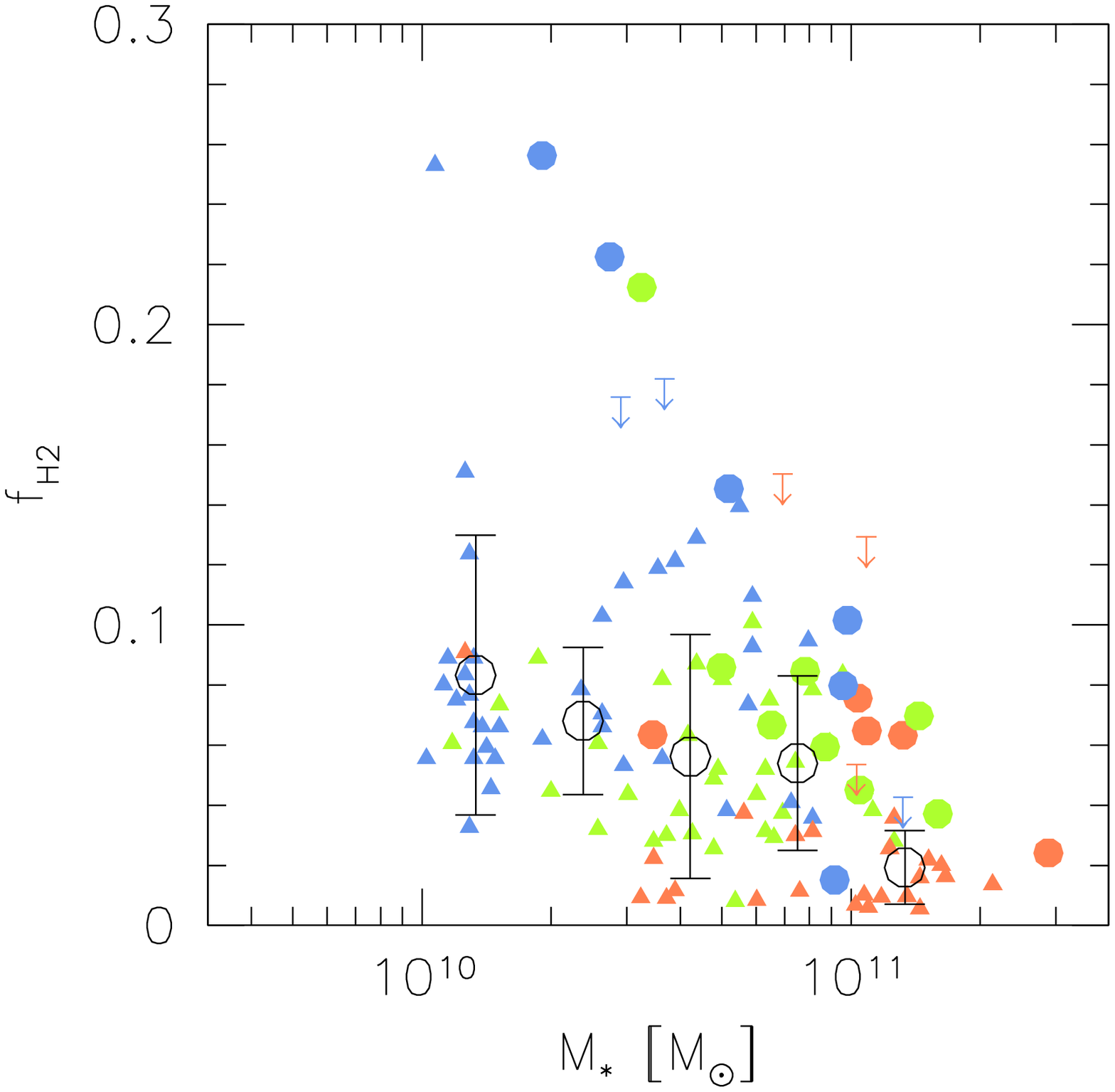,width=0.34\linewidth,clip=} 
\vspace{-0.3cm}
 \caption{Total gas mass fraction  ($f_g \equiv \frac{M_{\rm HI}+M_{\rm H2}}{M_{\rm HI}+M_{\rm H2}+M_*}$) and molecular gas  mass fraction  ($f_{\rm H2} \equiv \frac{M_{\rm H2}}{M_{\rm H2}+M_*}$) as a function of stellar surface mass density ($\mu_*$, left panel) and stellar mass ($M_*$, middle and right panel).  Ensemble means and standard deviations for the COLD GASS sample are shown as large empty circles with error bars in the middle and right panel.  Symbols and colours are the same as Figure~\ref{fig_mhimh2}.}
 \label{fig_fgas}
\end{figure*}

\subsubsection{Molecular and total gas mass fraction}
\label{sec:fg}

Among the galaxies detected in both H{\sc i} and CO, the gas to stellar mass fraction of the H{\sc i} monsters, $M_{\rm HI+H2}/M_*=0.74\pm0.57$, is more than twice as large as  that of the { reference sample} ($0.35\pm0.31$). The difference grows to more than a factor of three when galaxies with \hi\ monster-like neutral gas content ($M_{\rm HI} \ge 10^{10} M_\odot$) are removed from the COLD GASS sample.  Cold gas is the dominant component of the baryonic mass budget in some of these galaxies, particularly among the \hi\ monsters.

Both molecular gas mass fraction and total gas mass fraction are important in understanding mass build-up history of galaxies.  \citet{mcgaugh97} have shown that total gas mass fraction of spiral galaxies is strongly correlated with luminosity and surface brightness and can be reproduced by a simple disk evolution model.  McGaugh \& de Blok acknowledged the importance of molecular gas as the main fuel for star formation, but they did not have the appropriate data to incorporate molecular gas into their analysis properly.  Instead they modeled the molecular gas content based on the observed trends between $R_{mol}$ and Hubble type \citep{young89}.  As shown in the previous section, \hi\ dominates the total cold gas content among local galaxies, and this model-based accounting of molecular gas contribution probably had only a minor impact on the conclusions by McGaugh \& de Blok.

The correlation between total gas mass fraction and surface brightness reported by McGaugh \& de Blok is reproduced on the left panel of Figure~\ref{fig_fgas}.  A broad trend of a higher gas mass fraction for galaxies with a lower stellar density is present although dispersion in the trend is significantly larger in the new data (compared with their Figure~7).  The difference may arise from our using actual \htwo\ masses derived from CO measurements and using $\mu_*$ rather than central $B$-band surface brightness.  Colour-tagging galaxies by their $u-r$ colour also reveals that much of the correlation reflects the segregation of galaxies with different colour (i.e., star formation history).  One clear outlier in this plot is the \hi\ monster AGC192040, which is an unusual galaxy with a compact, red central stellar component and ring-like stellar arms (see \S~\ref{stellar mass density and conc}).

Total gas mass fraction ($f_g \equiv \frac{M_{\rm HI}+M_{\rm H2}}{M_{\rm HI}+M_{\rm H2}+M_*}$) is compared with stellar mass $M_*$ in the middle panel of Figure~\ref{fig_fgas}.  Two trends are obvious: (1) total gas mass fraction increases rapidly with decreasing stellar mass; and (2) total gas mass fraction of \hi\ monsters are nearly twice as large as the ensemble average of the COLD GASS sample at a given stellar mass.  Both of these trends could have been predicted from Figures~\ref{fig_hipro}~\&~\ref{fig_h2pro} which showed that \hi\ and \htwo\ masses are independent of $M_*$ and that \hi\ monsters have the highest $M_{\rm HI}$ and $M_{\rm H2}$ at a given stellar mass.  The \hi\ monsters trace the upper envelop of the gas mass fraction distribution, well above the ensemble mean trend.  In fact, most of the \hi\ monsters are more than $2\sigma$ above the ensemble mean, suggesting that they may represent a distinct population.

Molecular gas mass fraction ($f_{\rm H2} \equiv \frac{M_{\rm H2}}{M_{\rm H2}+M_*}$), shown on the right panel of Figure~\ref{fig_fgas}, display a rather different behavior from the total gas mass fraction just discussed.  Unlike the total gas mass fraction $f_g$, molecular gas mass fraction $f_{\rm H2}$ is a much flatter function of stellar mass, with only a weak systematic trend with $M_*$ is seen. Furthermore, \hi\ monsters mostly {\em follow} the same trend as the COLD GASS sample, rather than standing apart as did for the total gas mass fraction as discussed above, although a few exceptions stand out.  This completely different behavior was already seen in the comparison of $\mu_{\rm HI}$ and $\mu_{\rm H2}$ with $\mu_*$ in Figure~\ref{fig_sigmas}, which suggests a close physical link between stellar mass and molecular gas and a disconnect from neutral atomic gas.  The nearly linear correlation seen between $\mu_{\rm H2}$ and $\mu_*$ would naturally lead to a flat dependence of $f_{\rm H2}$ on $M_*$ with a mean ratio of 0.05-0.08 across the entire stellar mass range representing late type galaxies.  

\subsection{Implications on modeling galaxy gas content and star formation rate evolution}
\label{sec:theory}

Obtaining better understanding of gas physics and cold gas content evolution is a critical next step in connecting galaxy evolution theory with observations of baryonic tracers \citep{yun09,putman09,carilli13}, and this realization has spun a series of recent papers on modeling cold gas content in galaxies and its cosmic evolution \citep{or09,bau10,power10,lagos11a,lagos11}.
The weak correlations we found among \hi, \htwo, and stellar mass for the \hi\ monsters and the COLD GASS comparison sample bode poorly for semi-analytic models that incorporate mainly the mean global trends.  For example,  \citet{lagos11} have shown that modeling gas content through post-processing of a cosmological simulation \citep[e.g.,][]{or09,power10} does a poor job of recovering the local \hi\ mass function, while a self-consistent approach building on the derived merger trees is more successful.  A series of tests performed by \citet{lagos11a,lagos11} have also shown that different star formation recipes can lead to drastically different gas masses and molecular fractions, demonstrating the necessity for a certain level of sophistication and care in modeling gas physics.  

An illuminating lesson is the finding by \citet{lagos11a} that a star formation recipe with a linear dependence on gas density ($\Sigma_{SFR}\propto\Sigma_{gas}$) does a better job of reproducing the observed gas content in local galaxies than a non-linear formulation (e.g., Schmidt-Kennicutt ``law", $\Sigma_{SFR}\propto\Sigma_{gas}^N$).  While ample evidence exists for a non-linear dependence of $\Sigma_{SFR}$ on $\Sigma_{gas}$ on sub-kpc scales \citep[e.g., ][]{bigiel08}, we have shown in Figure~\ref{fig_sigmas} that the observed {\em global} dependence is closer to linear, with an additional complication of a colour-dependence.    This empirical trend, as well as the tight linear trend between $M_{\rm H2}$ and SFR discussed in \S~\ref{sec:sfr}, suggests that a recipe with a linear dependence should be more effective, regardless of the true underlying physics, when applied on global scales.  

Another important clue for future modeling studies revealed by our analysis of cold gas content is the clear disconnect between atomic and molecular gas, particularly in relation to stellar mass and SF activities. The strong dependence of total gas mass fraction $f_g$ on stellar mass $M_*$ seen in the middle panel of Figure~\ref{fig_fgas} offers a stark contrast to the nearly constant molecular mass fraction $f_{\rm H2}$ as a function of $M_*$ seen in the right panel.  The absence of correlation between {\em globally averaged} $\mu_{\rm HI}$ with $\mu_\ast$ also indicates the disconnect in spatial distribution between atomic gas and stellar disk, as well as between total \hi\ and stellar mass, as discussed in \S~\ref{CO measurement}. This contrasting behavior between atomic and molecular gas demonstrates a clear need to track the gas within the stellar disk separately from the gas in the larger halo.  %Few recent works, including those by Lagos et al., have yet to incorporate a method to address this need for different time and spatial scales into their models.  

A disproportional attention given to molecular mass fraction, motivated by the general expectation of a higher $f_{\rm H2}$ associated with a higher star formation rate in galaxies in the earlier epochs \citep[e.g.,][]{daddi10,tacconi10,geach11}, may also be misplaced.  While  \htwo\ mass clearly correlates closely with far-$IR$ luminosity and {\em current} SFR, \htwo\ is a minor component in the total cold gas content and accounts for less than 10\% of the baryon mass budget in \hi\ monsters and COLD GASS sample.  Also seen in Figure~\ref{fig_fgas} is that \hi\ dominates the entire baryon mass budget among some of the galaxies with stellar mass up to $5\times10^{10}M_\odot$, including many of the \hi\ monsters.  Focussing on the total baryon mass budget, \citet{mcgaugh97} have shown that {\em total} gas mass fraction is closely tied with the integrated star formation history among disk galaxies, and this implies total cold gas content (\hi+\htwo) is the important parameter in understanding the evolution of galaxies and stellar mass build-up.  We confirm the same general trend using a larger sample with a wider range of galaxy types with actual molecular gas measurements.  Furthermore, we find the {\em total} gas depletion time to match or to exceed the Hubble time among nearly all \hi\ monsters and a large fraction of COLD GASS galaxies detected in \hi\ and CO.  It is widely believed that \hi\ content and total cold gas budget are regulated by gas accretion from the surrounding halo and quenching induced by environment.  Future modeling studies should examine the effectiveness of the recipes dealing with halo gas accretion and quenching and the role of the environment in this regard.  While little 21cm \hi\ data exists for galaxies beyond $z\sim 0.1$ today to offer a useful test for these studies, ongoing studies such as the Blind Ultra Deep H I Environmental Survey \citep[BUDHIES, ][]{verheijen07,jaffe12} and the COSMOS \hi\ Large Extragalactic Survey \citep[CHILES, ][]{ximena13} should begin to produce blind surveys of cold gas content covering cosmologically interesting volume to $z\sim 0.45$.  

\section{Summary}
\label{summary}
In this work, we investigate molecular gas properties of galaxies with massive atomic gas reservoirs called H{\sc i} monsters, using complementary SDSS DR7 and MPA-JHU DR7 database as well as the WISE 22~\micron\ photometry.  Selected only for their \hi\ mass ($M_{\rm HI} \ge 2\times 10^{10}~M_{\odot}$), these galaxies provide us with a unique opportunity to study how such gas-massive systems are distinguished in their physical properties (star formation activity, molecular gas content, stellar mass and colour, etc) from normal massive galaxies ($M_{\ast} \ge 10^{10}~M_{\odot}$) such as those recently studied by the COLD GASS team \citep{as1}.%, possibly yielding important new insights on the cold gas content evolution and star formation history.

Among the 28 \hi\ monsters observed, 19 galaxies are detected in CO with H$_{2}$ masses of $(1 - 11)\times10^{9}~M_{\odot}$, derived using a standard ${\rm CO} - {\rm H_{2}}$ conversion factor.  The median H$_2$ mass, $7.2\times10^9~M_{\odot}$, is 50\% larger than that of the $L^*$ galaxy of the local CO luminosity function \citep{k03}. Therefore, H{\sc i} monsters represent some of the most {\em molecular gas-massive} galaxies in the Local Universe, independent of their stellar properties. 
$M_{\rm HI}$ and $M_{\rm H2}$ are broadly correlated so that the more H{\sc i} rich galaxies also tend to have more H$_2$, but the scatter in the relation is considerable. The H{\sc i} monsters lie at the upper end of the $M_{\rm HI}$ and $M_{\rm H2}$ mass spectrum, extending the trend exhibited by COLD GASS galaxies \citep{as1}.  This general trend supports the idea that molecular hydrogen forms out of neutral atomic hydrogen, but the observed large scatter indicates that the situation is more complex.  The molecular-to-atomic gas mass ratios for the \hi\ monsters are similar to those of the COLD GASS sample, which has the mean of $<M_{\rm H2}/M_{\rm HI}> \approx 0.3$ \citep{as1} but with a factor of $\sim10$ scatter.   This dispersion is significantly reduced when the two gas masses are normalized by their stellar mass.  This analysis also shows that stellar  $u-r$ colour is another major component leading to the spread in this relation.  

By analyzing the gas and stellar properties of the \hi\ monsters and the COLD GASS sample together, we confirm the findings by \citet{as1, as2} that the $u-r$ colour is tightly correlated with molecular gas fraction ($M_{\rm H2}/M_{\ast}$) while the other stellar properties are only weakly correlated at best.  The global fractional molecular gas mass to that of atomic gas ($R_{mol} \equiv M_{\rm H2}/M_{\rm HI}$) shows little correlation with stellar mass density $\mu_*$ when both samples are examined together, contrary to the tight correlation seen in spatially resolved distribution in nearby spiral disks \citep{br06}.  
Our detailed analysis shows that $\mu_{\rm H2}$ and $\mu_*$ are indeed tightly correlated, especially when the dependence on $u-r$ colour is also taken into account.  On the other hand, $\mu_{\rm HI}$ and $\mu_*$ are completely uncorrelated on global scales.  This disconnect between \hi\ and stellar component on global scales is not surprising since \hi\ disks typically extend far beyond their stellar counterparts \citep[e.g.,][]{walter08}.  The apparent correlation between $R_{mol}$ and $\mu_*$ among the \hi\ monsters reflects the underlying correlation between $\mu_{H2}$ and $\mu_*$ because the \hi\ monsters have only a narrow range of $M_{\rm HI}$.  

We found a broad correlation between molecular mass fraction ($M_{\rm H2}/M_*$) and SFR among all galaxies, and the \hi\ monsters are indistinguishable from the COLD GASS sample in this trend.  An even tighter correlation is seen between $M_{\rm H2}/M_*$ and sSFR~(=SFR/$M_*$), and we identify this tighter trend as that of the well-known near-linear correlation between $M_{\rm H2}$ and SFR.  In this comparison, the \hi\ monsters stand out again with a systematically {\em lower} star formation efficiency.  The cause for this curiously lower SF efficiency for the \hi\ monsters is unknown, and understanding this would require additional future work including the analysis of spatially resolved gas and star formation distributions.  The gas depletion time exceeds the Hubble time for many \hi\ monsters and COLD GASS sample galaxies, suggesting that the need for continuous gas accretion from the more extended hot halo may not be as critical as previously thought \citep[e.g.,][]{putman06}, at least for these late type galaxies containing significant amount of cold gas already.  The total cold gas mass fraction shows a clear trend with stellar mass and colour, suggesting a strong link between star formation history and total gas content, similar to the scenario previously suggested by \citet{mcgaugh97} and others.   In contrast, molecular gas mass fraction is nearly constant over the whole range of stellar mass and colour, indicating that the relevant time scales for the atomic and molecular gas are rather different and distinct.

Based on our analysis of atomic and molecular gas content among the \hi\ monsters and the COLD GASS sample, we offer a few cautions and suggestions for future modeling studies of cold gas content evolution.  First and foremost, we find clear evidence for spatial and temporal disconnect between \hi\ and \htwo\ (and stars), and the modeling of these different phases of cold gas should be de-coupled, rather than linked together using a simple recipe.  When modeling global star formation rate and gas content evolution using a numerical simulation with a coarse spatial and time resolution, a star formation recipe with a linear dependence on $M_{\rm H2}$, as seen in our data, may be more effective than a more complex, non-linear recipe (e.g., ``Schmidt-Kennicutt law''). In general, adopting empirically driven recipes supported by observational data may be more effective in modeling the observations, rather than implementing more physically motivated micro-physics that are modified to fit over the spatial and temporal resolution limits.

\section*{Acknowledgments}
We are grateful to Mike Brewer, Don Lydon, Kamal Souccar, Gary Wallace, Ron Grosslein, John Wielgus, Vern Fath, and Ronna Erickson for their technical support of the Redshift Search Receiver commissioning.  We also thank Martha Haynes, Riccardo Giovanelli, and other members of the Arecibo Legacy Fast ALFA Survey (ALFALFA) team who contributed to the initial selection of the \hi\ monster sample from the early ALFALFA catalog and Jonathan T. Yun for contributing to the compilation of the WISE photometry from the NASA/IPAC Infrared Science Archive (IRSA). A.C. acknowledges support from SRC Program (2010-0027910) and 2013-8-0579 of the National Research Foundation of Korea, and Science Fellowship of POSCO TJ Park Foundation. This work was also supported by NSF grants AST 0096854, AST 0540852, and AST 0704966. This research has made use of the NASA/ IPAC Infrared Science Archive, which is operated by the Jet Propulsion Laboratory, California Institute of Technology, under contract with the National Aeronautics and Space Administration. This publication makes use of data products from the Wide-field Infrared Survey Explorer, which is a joint project of the University of California, Los Angeles, and the Jet Propulsion Laboratory/California Institute of Technology, funded by the National Aeronautics and Space Administration. Funding for SDSS and SDSS-II has been provided by the Alfred P. Sloan Foundation, the Participating Institutions, the National Science Foundation, the U.S. Department of Energy, the National Aeronautics and Space Administra- tion, the Japanese Monbukagakusho, the Max Planck Soci- ety, and the Higher Education Funding Council for England. The SDSS web site is http : //www.sdss.org.

\label{lastpage}

\end{document}